\documentclass[fleqn,usenatbib]{mnras}
\usepackage[english]{babel}
\usepackage[T1]{fontenc}
\usepackage{ae,aecompl}
\usepackage{fancyhdr}
\usepackage{amsfonts}
\usepackage{amsmath}
\usepackage{amssymb}
\usepackage{multicol}
\usepackage{layout}
\usepackage{graphicx}
\usepackage{multirow}
\usepackage{color}
\usepackage{multirow}
\usepackage{times}
\usepackage{natbib}
\def\be{\begin{equation}}
\def\ee{\end{equation}}

\usepackage[utf8]{inputenc}
\usepackage{lmodern}

\definecolor{valecol}{rgb}{0,0.5, 1.}

\newif\ifAMStwofonts
\AMStwofontstrue

\graphicspath{{./fig/}}

\title[MOG cosmology without dark matter and  the cosmological constant ]{MOG cosmology without dark matter and the cosmological constant}

\author[Davari, Rahvar]{
	Zahra Davari$^{1}$ and Sohrab Rahvar$^1$ \\ 
	$^1$ Department of Physics,Sharif University of Technology, P.O.Box 11365-9161, Tehran,
	Iran\\}
\date{Accepted ?, Received ?; in original form \today}

\pagerange{\pageref{firstpage}--\pageref{lastpage}} \pubyear{2015}

\begin{document}
	\label{firstpage}
	
	\maketitle	
	\begin{abstract}	
In this work, we investigate the MOdified Gravity (MOG) theory for dynamics of the universe and compare the results with the $\Lambda$CDM cosmology. We study the background cosmological properties of the MOG model and structure formation at the linear perturbation level. We compare the two models with the currently available cosmological data by using statistical Bayesian analyses. After obtaining updated constraints on the free parameters, we use some methods of model selection to assist in choosing the more consistent model such as the reduced chi-squared ($\chi^2_{\rm red}$) and a number of the basic information criteria such as the Akaike Information Criterion (AIC), the Bayes factor or Bayesian Information Criterion (BIC), and Deviance Information Criterion (DIC). MOG model appears to be consistent with the $\Lambda$CDM model by the results of $\chi^2_{\rm red}$ and DIC for an overall statistical analysis using the background data and the linear growth of structure formation.
	\end{abstract}
	
	
	
\section{Introduction} \label{sec:intro}
The impressive and wonderful success of the standard $\Lambda$CDM model across many different scales and in a wide range of cosmological observations indicate that $\Lambda$CDM is superior to other models due to its simplicity and fewer free parameters. 
 However, this standard model involves many observational problems in cosmology, such as the lack of the nature of dark matter particles (DM) as well as the dark energy (DE). 
 Moreover, recent advances in precision cosmology have yielded discrepancies in observations at different redshifts, in particular, the following tensions in the $\Lambda$CDM model: $H_0$ tension~\citep{Aghanim:2018oex,Riess:2019cxk}, $S_8$ tension~\citep{Kumar:2019wfs}, Cosmic Microwave Background (CMB) high-low l tension~\citep{Hinshaw:2012aka}, Baryon Acoustic Oscillation (BAO) ly-$\alpha$ tension~\citep{McDonald:2004xn,Slosar:2013fi}, the suppress of the high angular scale correlation in CMB power spectrum ~\citep{Bernui:2018wef}, signature for the violation of the statistical isotropy of CMB ~\citep{Schwarz:2015cma}, which have opened the door to several alternative models. According to the $\Lambda$CDM model, only about $5\%$ of the universe is either visible or detectable.
This fact has become a powerful motivator for the search for alternative explanations that can explain cosmic observations without regard to dark matter or cosmological constant or both of them.\\
One of these alternative models is the fully covariant and relativistic modified gravity(MOG) theory~\citep{Moffat:2005si}.
The MOG theory or Scalar-Tensor-Vector-Gravity (STVG) has been explained  successfully the rotation curves of galaxies and account for galaxy cluster masses\citep{rahvar1,rahvar2,Moffat:2014pia, Brownstein:2005zz,Green:2019cqm, Davari:2020ijn}
, velocity dispersions of satellite galaxies, globular clusters~\citep{Moffat:2020nmq}, and the Bullet Cluster without resorting to cold dark matter~\citep{Israel:2016qsf}. It is also shown that MOG theory satisfies the weak equivalence principle. Also this theory is in agreement with gravitational wave event of GRB170817A and corresponding gamma ray burster event of GRB170817A which could be explained as the merge of the neutron stars ~\citep{Green:2017qcv}.\\
The action of the MOG theory, in addition to Einstein-Hilbert and matter terms, includes a massive vector field and three scaler fields corresponding to running values of the gravitational constant, the vector field coupling constant, and the mass of vector field.\\
In this paper, we investigate the dynamics of the universe using MOG taking into account baryonic matter and radiation as the Energy- Momentum parameters of the universe. Also, we investigate linear structure formation in this model. \\
This paper is structured as follows: In section~(\ref{sec:1}), we first introduce the MOG theory and then review the key features of the background evolution of the universe in the MOG theory.
In section~(\ref{sec:2}), after that, we derive the main equations governing the evolution of baryonic matter in the linear perturbation and investigate the growth of matter perturbations in the MOG model.
We introduce the available cosmological data at background and perturbation levels and we perform a numerical Markov Chain Monte Carlo (MCMC) analysis in order to constraint the free parameters of the models and compare them to each other in sections~(\ref{sec:3}) and~(\ref{sec:4}). After that, we deal with the evolution of a number of some main cosmological quantities with the best results obtained with the data in the section~(\ref{sec:5}).
Finally, our findings are drawn in section~(\ref{sec:con}).
\section{Background evoluation in the MOG cosmology}\label{sec:1}
In order to study and evaluate the MOG theory at the background level, it is necessary to first obtain the main equations governing the evolution of the cosmic background within the MOG theory. 

The MOG theory consists of a massive vector field $\phi_\mu$, three scalar fields of (i) G, the coupling strength of gravity which can also be considered as a scalar field, (ii) $\omega$, which describes the coupling strength between the vector field and matter and (iii) $\mu$, that related  to the mass of the vector field ~\citep{Moffat:2005si, Moffat:2007nj}. The MOG action which includes scalar, vector and tensor fields is given by $S=S_\phi+S_G+S_S+S_M$, that
	\begin{eqnarray}
		&&S_G=-\frac{1}{16\pi}\int \frac{1}{G}(R+2\Lambda)\sqrt{-g}d^4x,\\
		&&S_{\phi}=-\frac{1}{4\pi}\omega \int [\frac{1}{4}B^{\mu\nu}B_{\mu\nu}-\frac{1}{2}\mu^2\phi_{\mu}\phi^{\mu}+V_{\phi}(\phi)]\sqrt{-g}d^4x, \nonumber\\
		&&S_S=-\int\frac{1}{G}[\frac{1}{2}g^{\alpha \beta}(\frac{\nabla_{\alpha}G\nabla_{\beta}G}{G^2}+\frac{\nabla_{\alpha}\mu\nabla_{\beta}\mu}{\mu^2})+\nonumber\\
		&&\qquad\qquad\qquad\frac{V_G(G)}{G^2}+\frac{V_{\mu}(\mu)}{\mu^2}]\sqrt{-g}d^4x,\nonumber
	\end{eqnarray}
where $S_M$ is the matter action, $B_{\mu\nu}=\partial_{\mu}\phi_{\nu}-\partial_{\nu}\phi_{\mu}$ is the Faraday tensor for the vector field and $V_{\phi}(\phi)$, $V_G(G)$ and $V_{\mu} (\mu)$ represent potentials corresponding to the vector field and the scalar fields~\citep{Carolina2020}.
The symbol $\nabla_{\nu}$ shows the covariant derivative with respect to the metric $g_{\mu\nu}$ and the symbols $R$ and $g$ are the Ricci-scalar and the determinant of the metric tensor, respectively.\\
We assume that the universe is isotropic and homogeneous. The observed isotropy of cosmic microwave background radiation is the main evidence that confirms this cosmological principle.  We base our cosmology on  Friedmann-Lemaitre- Robertson-Walker (FLRW) metric: 
	\begin{equation}\label{flrw}
		ds^2=dt^2-a^2(t)\left(\frac{dr^2}{1-kr^2}+r^2(d\theta^2+\sin^2\theta d\phi^2)\right),
	\end{equation}
where $a(t)$ is the scale factor and $k =0,-1, +1$ corresponds to flat, open and closed spatial geometry, respectively. Since observations point to almost spatial flatness, we shall restrict ourselves to $k = 0$ throughout the work. Due to the symmetries of the FLRW background spacetime, we set $\phi_i=0 (i = 1, 2, 3)$ and $B_{\mu\nu}=0$ and also in~\cite{Moffat:2007nj} is supposed the value of G changes sharply, although smoothly; meanwhile, the scalar fields $\omega$ and $\mu$ remain constant.
We use the energy-momentum tensor of a perfect fluid:
	\begin{equation}
		T^{\mu\nu}=(\rho+p)u^\mu u^\nu -pg^{\mu\nu},
	\end{equation}
where $u^\mu=dx^\mu/ds$ and $\rho$ and p are the density and pressure of matter. As it has discussed in section~(\ref{sec:intro}) in the MOG theory, we do not include dark matter and constant cosmological as dark energy to explain astrophysical and cosmological observations. So we have 
	\begin{equation}
		\rho=\rho_B+\rho_R+\rho_G+\rho_k+\rho_V,
	\end{equation}
where $\rho_B$ and $\rho_R$  denote the density of baryons and   radiation(photons and neutrinos) and $\rho_G$,$\rho_k$ and $\rho_V$ are the kinetic, the potential of the Brans-Dicke (the scalar field G) terms, respectively. They are further explained in the following. Since radiation and baryons do not interact with each other, thus they follow the standard equations of evolution i.e  $\rho_R(a)=\rho_{0r}a^{-4}$ and $\rho_B(a)=\rho_{0b}a^{-3}$, respectively.\\
The modified Einstein field equations for the MOG theory can be written  in an explained geometric structure of the universe follow as:~\citep{Moffat:2014bfa}:
	\begin{equation}\label{H}
		H^2+\frac{k}{a^2}=\frac{8\pi G}{3}\rho-\frac{4\pi}{3}(\frac{\dot{G}^2}{G^2})+\frac{8\pi}{3}(\frac{V_G}{G^2})+H\frac{\dot{G}}{G},
	\end{equation}
	\begin{equation}\label{Hp}
		\frac{\ddot{a}}{a}=-\frac{4\pi G}{3}(\rho+3p)+\frac{8\pi}{3}(\frac{\dot{G}^2}{G^2})+\frac{8\pi}{3}(\frac{V_G}{G^2})+H\frac{\dot{G}}{2G}+\frac{\ddot{G}}{2G}-\frac{\dot{G}^2}{G^2},
	\end{equation}
where an overhead denotes derivative with respect to the cosmic time and $H \equiv \dot{a}/a$ is the Hubble parameter. On the other hand, the equation of motion for the scalar field G obtain as:
	\begin{equation}\label{G}
		\ddot{G}+3H \dot{G}-\frac{3}{2}\frac{\dot{G}^2}{G}+\frac{3}{G}V_G-V'_G-\frac{3G}{8\pi}\left(\frac{\ddot{a}}{a}+H^2\right)=0,
	\end{equation} 
where $V'_G=dV_G/dG$. We can rewrite the Friedmann equations into the familiar form:
\begin{equation}\label{Hrho}
H^2=\frac{8\pi G}{3}\sum_i \rho_i=\frac{8\pi G}{3}[\rho+\rho_V+\rho_k+\rho_G],
\end{equation}
	and  
	\begin{align}\label{Hp2}
	\frac{\ddot{a}}{a}&=-\frac{4\pi G}{3}[\rho(1+3w)+\rho_V(1+3w_V)+\rho_k(1+3w_k)\\ \nonumber
	&+\rho_G(1+3w_G)]-\frac{\dot{G}^2}{G^2}+\frac{\ddot{G}}{2G},
	\end{align}
where  $\rho_V=\frac{1}{G}(\frac{V_G}{G^2})$, $\rho_k=-\frac{1}{2G}(\frac{\dot{G}2}{G^2})$ and $\rho_G=\frac{3H\dot{G}}{8\pi G^2}$ are the potential, the kinetic and Brans-Dicke terms, respectively.
The equations of state (EoS) associated with $\rho_k$, $\rho_V$, and  $\rho_G$ by comparing the equations~(\ref{Hp2}) and (\ref{Hp}) are obtained 1,-1 and $-2/3$, respectively~\citep{Moffat:2007ju}.\\
It has been shown that for the proper description of cosmological quantities such as evolutionary stages of the universe, the age and  the redshift space distortion (RSD) data, $V_G$ must be different from 0, in addition, in previous studies of the MOG theory had considered $V_G$ as a constant~\citep{Toth:2010ah, Jamali:2018uij}. 
In this work, we will consider it as a power-law function inspired by the original ~\cite{Peebles1988} potential: $V_G=V_{G0} G^{-\beta}$ where $V_{G0}$ and $\beta$ are free parameters.\\
It should be noted that another example of  Scalar-Tensor  gravity  theories is Brans-Dicke theory (BD). The gravitational constant, G in Brans-Dicke theory  assumed as an inverse of the scalar field $\phi$. The field equation in BD theory contains a dimensionless parameter, $\omega_{BD}$, called the Brans-Dicke coupling constant and it can be determined by fitting to the astronomical observations and astrophysical experiments. There are similarities between BD theory and the MOG theory that we have considered but the equations (5) to (7) are different from the relations obtained for Brans-Dicke theory in references ~\cite{Hrycyna:2014cka, Sola:2020lba} and  \cite{Perez:2021kdh} since we assume a potential function with a free coefficient, $\beta$ that it is different from similar works in Brans-Dicke theory. Also here, we do not consider both the dark matter and the cosmological constant in the content of the universe.\\
It is appropriate to represent equations in terms of the scale factor using $\frac{d}{dt}=aH\frac{d}{da}$ for example we can rewrite the equation \eqref{H} as 
	\begin{equation}
		H(a)=\sqrt{\frac {H_0^2 \tilde{G}(a)[\Omega_{b0}a^{-3}+\Omega_{r0}a^{-4}]+\frac{8\pi }{3}\tilde{V}_{G0} \tilde{G}(a)^{-\beta-2}}{1+\frac{4\pi}{3}a^2\frac{\tilde{G'}^2(a)}{\tilde{G}^2(a)}-a\frac{\tilde{G'}(a)}{\tilde{G}(a)}}};
	\end{equation} 
  the current critical energy density are defined as $\rho_{c0}=3H_0^2/8\pi G(1)$ and the present values of the cosmological density parameters $\Omega_i=\rho_i/\rho_{c0}$, we also define $\tilde{G}(a)=G(a)/G(1)$ and $\tilde{V}_{G0}=V_{G0}G(1)^{-\beta-1}$.
In order to find the dynamics of G(a), we solve numerically in the equation~(\ref{G}) with suitable initial conditions. In most previous literature,  the initial conditions were considered as follows in the present time ($a=1$): $G(1)=G_N$ and $G'(1)=0$. In this work, we take on an ansatz of the power-law  to determine the initial conditions in the early universe:
	\begin{equation}\label{Gt}
		a=a_0 t^{n_1},\qquad \tilde{G}=G_0 t^{n_2};
	\end{equation}
where t is normalized to $t_0$(the age of the universe) in the present time 
then for matter dominated epoch we have zero pressure$(P=0)$ and $\rho\propto a^{-3}\propto t^{-3n_1}$. Substituting these relations in  the field equation leads to
	\begin{equation}
		n_1=\frac{2(\beta+3)}{3(\beta+2)}, \qquad n_2=\frac{2}{\beta+2};
	\end{equation}
	and
	\begin{equation}
		G_0=\left[\frac{12\pi \tilde{V}_{G_0} (\beta+2)^2}{(6-24\pi+\beta)}\right]^{\frac{1}{\beta+2}},
	\end{equation}
where we can constrain the parameters of model with the numeric value of $G_0$. For radiation dominated epoch, we obtain $n_1=\frac{(\beta+3)}{2(\beta+2)}$.  
In the GR Einstein- de Sitter universe model (EdS), for matter or radiation, we have $a\propto t^{2/3}$ or $a\propto t^{1/2}$ but here we obtain  $a\propto t^{2(\beta+3)/3(\beta+2)}$ or $a\propto t^{2(\beta+3)/4(\beta+2)}$ for matter and radiation dominated epoch, respectively where the MOG model behavior similar to EdS model for $\beta\rightarrow\infty$.
We can substitute the cosmic time in \eqref{Gt} in terms of the scale factor. We use $H=n_1/t$ and can obtain $t^2=3 n_1^2/8\pi G_0\rho_{B}$ for the matter dominated epoch where $\rho_B=\rho_{c0}\Omega_{b0}a^{-3}$. On the other hand, with  an effective gravitational constant of $G_ {\rm eff} \approx 6G_N$, the effective baryonic matter taking into account the factor of six and contribute$\sim30\%$ of the critical density of the universe, therefore no nonbaryonic dark matter is needed to explain a deficiency in matter density. We use $t^2=n_1^2 a^3/6 H_0^2 \Omega_{b0}$ and determine $\tilde{G}$ as a function of the scale factor as
\begin{equation}\label{eqini}
		\tilde{G}(a)=\left(\frac{8\pi \tilde{V}_{G_0} (\beta+3)^2}{9\times 10^4 h^2 \Omega_{b0}(6-24\pi+\beta)}\right)^{\frac{1}{\beta+2}}a^{\frac{3}{\beta+3}},
\end{equation}
where h, the reduced Hubble constant is usually defined according to $H_0\equiv100 h$. As $\tilde{V}_{G_0}>0$, the equation \eqref{eqini} imposes $\beta>24\pi-6\simeq 69.4$. 
	\begin{figure}
		\begin{center}
			\includegraphics[width=8cm]{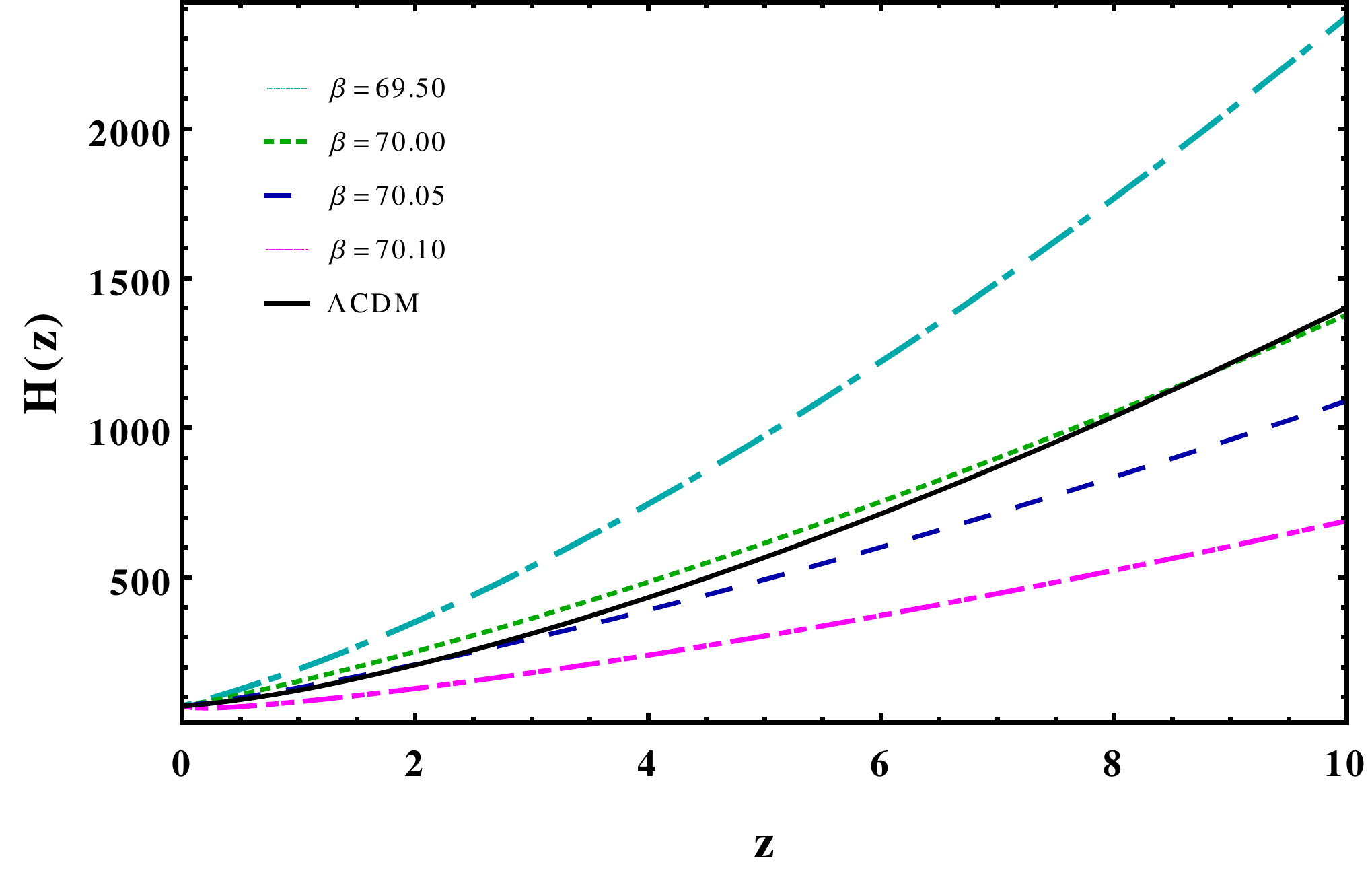}				\includegraphics[width=8cm]{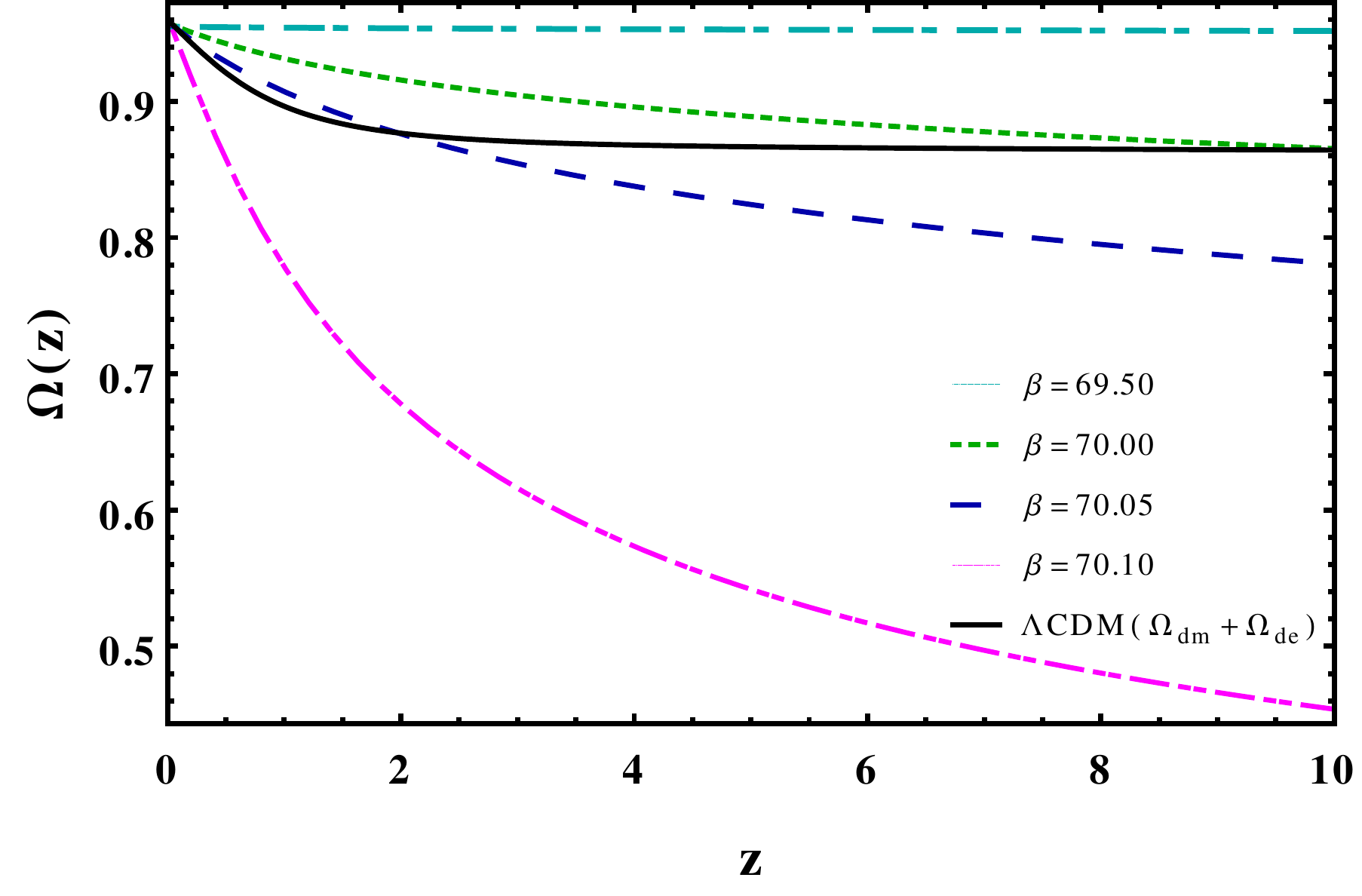}
			\caption{The redshift evolution of $H(z)$ (top panel), $\Omega(z)$ (bottom panel).
				The long dashed, dotted, dashed and dot-dashed curves correspond to MOG models with $\beta=69.5,70.0,70.05$ and $70.10$, respectively. We show the concordance $\Lambda$CDM model with the black solid curve.
.}
			\label{fig1}
		\end{center}
	\end{figure} 
In the following, we survey some of the most important parameters describing the expansion of the universe such as the present value of the Hubble parameter known as Hubble constant and the fraction $\Omega=\rho/\rho_c$, where $\rho_c$ is the critical density that corresponds to a flat universe.
We survey the behavior of the Hubble parameter in MOG cosmology since the Hubble expansion affects the growth of matter perturbations. In Figure~(\ref{fig1}), we show the redshift evolution of the Hubble parameter, $H(z)$ (top panel) for different values of the MOG parameter $\beta=69.50$ (long dashed cyan line), $\beta=70$ (dotted green line), $\beta=70.05$ (dashed blue line) and  $\beta=70.10$ (dot-dashed pink line). We also plotted the concordance $\Lambda$CDM model for comparison( the solid black line). As expected, it depends on the choice of $\beta$ and we see that in the case of $\beta<70$, we have $H_{MOG} (z) > H_{\Lambda} (z)$, while the opposite holds for $\beta>70$ and here the MOG has the best compatibility for $\beta=70.05$ with observational data and for $\beta=70$ has a similar behavior of the standard model. \\
In bottom panel of Figure~(\ref{fig1}), we plot the evolution of the energy density parameter of the scalar field, $\Omega_V(z)$ for all aforementioned values of $\beta$. As mentioned it played both dark matter and dark energy role. The panel shows $\Omega_V(z)$ for $\beta=70$ has similar behavior to $\Omega(z)$ in the $\Lambda$CDM case in large value of redshift. It should be noted that in these two panels of Figure~(\ref{fig1}) we adopted for MOG models $(\Omega_{b0}=0.04,H_0=70,V_{G0}=0.70)$ and also for the $\Lambda$CDM $(\Omega_{b0}=0.04,\Omega_{dm0}=0.26,H_0=70)$.\\
To further investigate the properties of the MOG theory, we define two
additional parameters: the effective equation of state parameter, $w_{\rm eff}$ and the deceleration parameter, q. $w_{\rm eff}$   is defined to include all components of the energy budget of the cosmos into an effective density$\rho_{\rm eff}$ and	effective pressure $p_{\rm eff}=w_{\rm eff}\rho_{\rm eff}$, such that the Friedmann equations for $\Lambda$CDM model is written as $H^2=8\pi G/3 \rho_{\rm eff}$ and $\ddot{a}/a=-4\pi G/3 (1+3w_{\rm eff})\rho_{\rm eff}$. It is given namely: $w_{\rm eff}=\frac{p_{\rm tot}}{\rho_{\rm tot}}=-1-\frac{2\dot{H}}{3H^2}$. Using the equations~(\ref{Hrho}) and (\ref{Hp}), we can write for MOG theory:
\begin{equation}\label{weff}
w_{\rm eff}=-1-\frac{2\dot{H}}{3H^2}-\frac{2(-\ddot{G}/2G+\dot{G}^2/G^2)}{3H^2}.
\end{equation}
In the top panel of Figure~(\ref{fig2}), we plot the evolution of the $w_{\rm eff}$ as a function of z. As expected, at the time of the current accelerated expansion $w_{\rm eff}$ at present time obtain $-0.02,-0.31,-0.57$ and $-1.85$ for $\beta=69.5,70.0,70.05$ and $70.10$, respectively and for $\Lambda$CDM is $-0.70$. We observe that for $\beta>70.08$, we reach the phantom regime prior to the present time ($w<-1$). In the case of $\beta=70.10$ the EoS parameter crosses the phantom line $w=-1$ at the epoch of $z \sim 0.2$, while for $\beta<70.10$ it  remains in quintessence regime for all z. \\
If we compare this panel with H(z) parameter in Figure~(\ref{fig1}) we observe for the quintessence MOG models ($\beta<70.08$), $H_{MOG}(z)>H_\Lambda$ for all redshifts which means that the corresponding cosmic expansion is larger than that of the concordance $\Lambda$CDM model.
Finally as a complementary information, in the bottom panel of the Figure~(\ref{fig2}), we present the evolution of a deceleration parameter $q=-\frac{\ddot{a}}{aH^2}=-1-\frac{\dot{H}}{H^2}$ in term of the redshift.
Recall that q for determining the accelerated phase of the expansion (q < 0) or decelerated phase (q > 0) can be used and $q = 0$ indicates the position of the transition point from decelerated expansion to accelerated expansion in the universe.
As expected, at the time of the current accelerated expansion, q is negative but we see in this panel only for two cases of MOG models that $\beta>70$, q are negative. The transition redshift from deceleration to acceleration for $\beta=70.05$ is at $z_t = 0.68$, which is in good agreement with the values obtained  for the $\Lambda$CDM ($z_t = 0.67$). Also q tends to $\frac{1}{2}$ that representing the deceleration phase in the matter dominated epoch. From the top panel of Figure~(\ref{fig2}), we can see for MOG model that for large redshifts $w_{\rm eff}$ approach into small negative value but not zero, unlike to $\Lambda$CDM model, however from the bottom panel of Figure~(\ref{fig2}) we  have deceleration phase for the large z (in the early universe). Figure~(\ref{fig2}) shows the fact that the effective equation of state and the deceleration parameter for both models are consistent and behave in a similar way.
These results are in good settlement with the values obtained in \citep{Farooq:2016zwm}.\\	
	\begin{figure}
		\begin{center}
			\includegraphics[width=8cm]{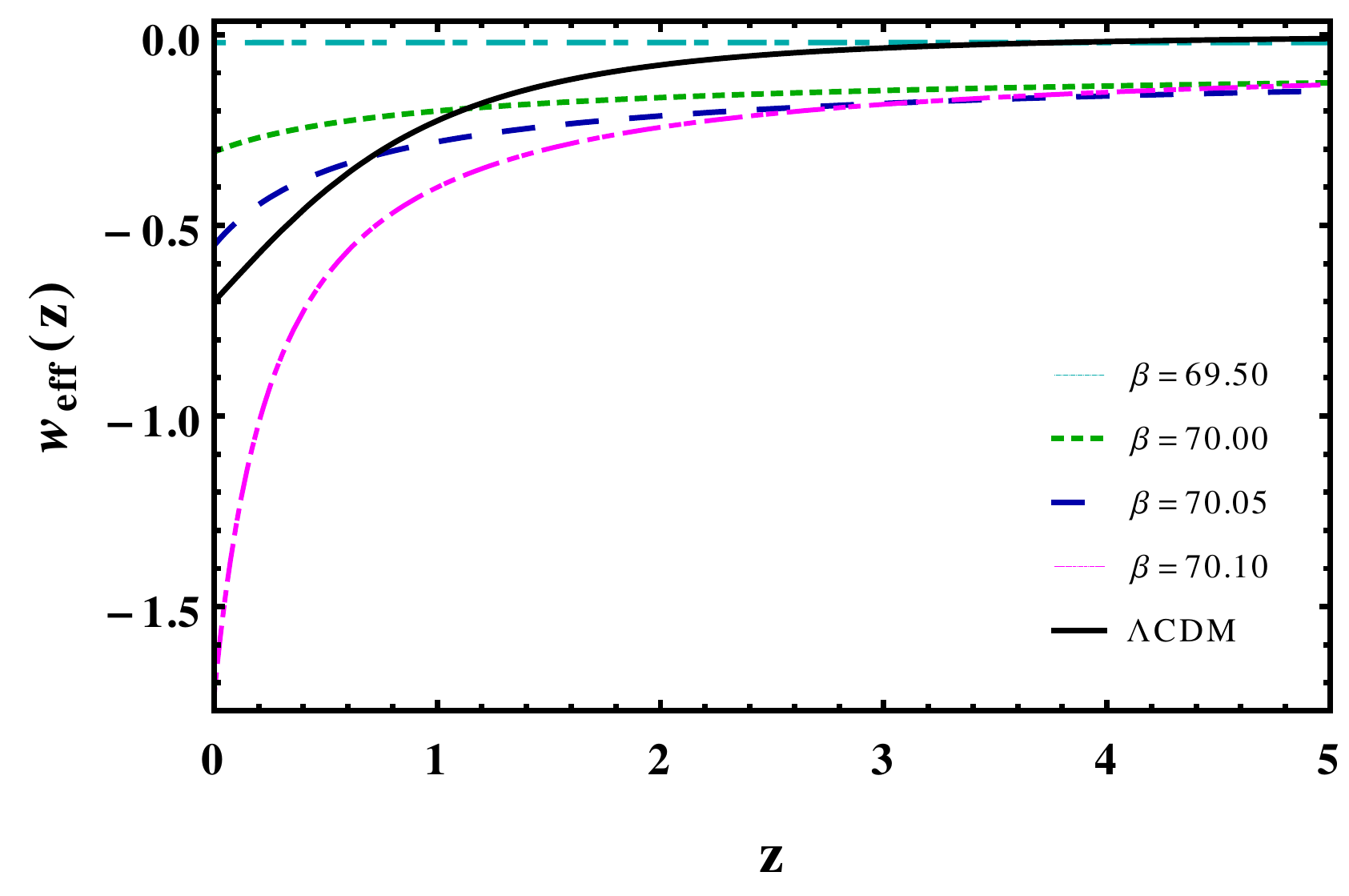}
			\includegraphics[width=8cm]{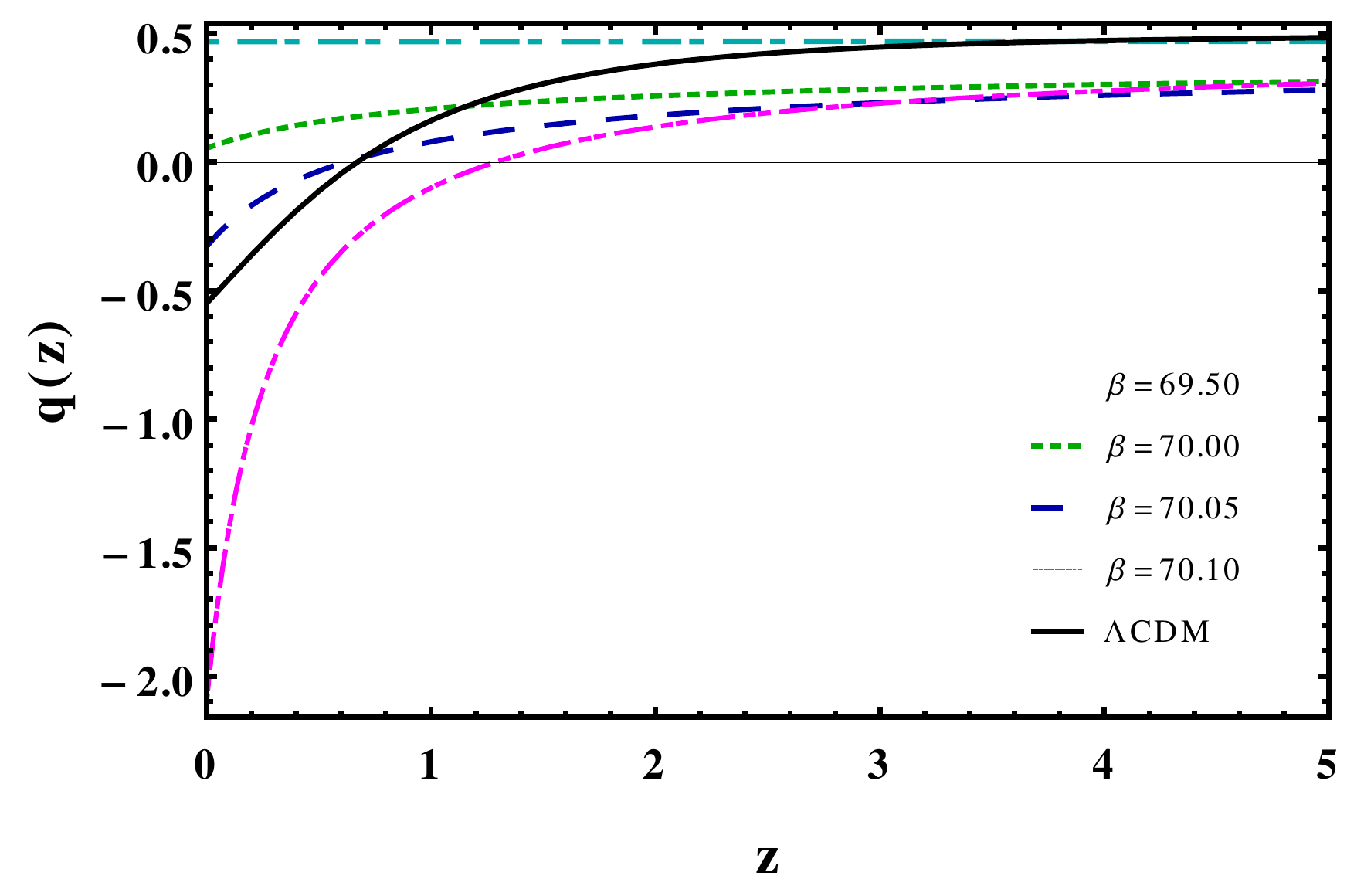}
									\caption{The redshift evolution of the deceleration parameter q(z) (top panel), the effective Equation of State, $\rm w_{\rm eff}$(bottom panel).
				The color of the curves used is similar to the Figure~(\ref{fig1}).}
			\label{fig2}
		\end{center}
	\end{figure} 
We show in Figure~(\ref{fig3}) G increase with scale factor for all values of free parameter $\beta$. In the following, we will present an argument based on simplified analytical calculations that MOG theory results in the accelerated expansion of the early universe similar to the standard period of cosmic inflation. It is obtained by substituting the relations~(\ref{Gt}) in the second Friedmann equation~(\ref{Hp}):
	\begin{equation}
		\frac{\ddot{a}}{a}=-\frac{6 + \beta - 48\pi}{9(\beta+2)^2},
\end{equation}
which is positive given the condition we have already obtained for the value
	$\beta>24\pi-6$.
	\begin{figure}
		\begin{center}
			\includegraphics[width=8cm]{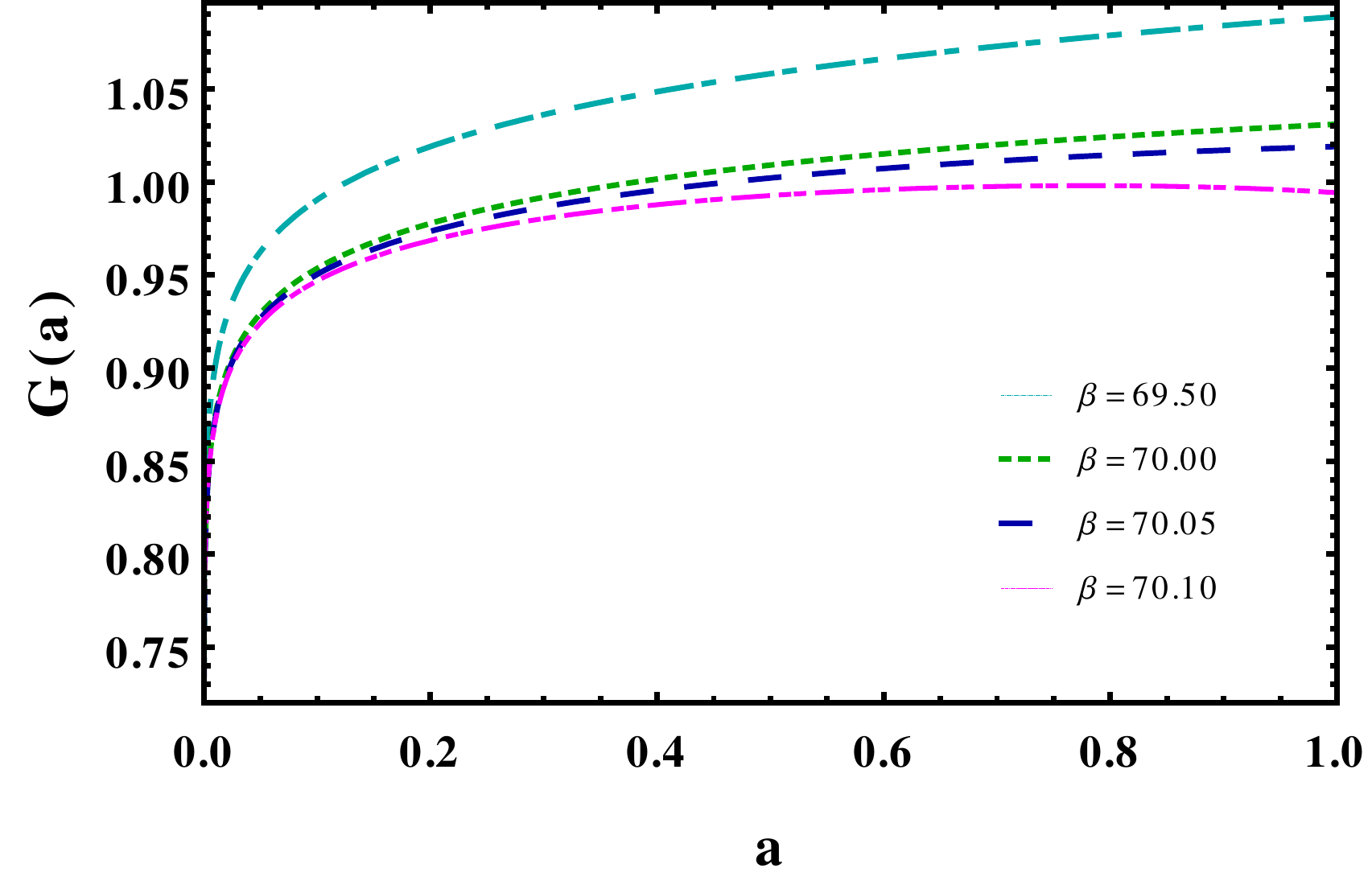}
			\caption{The evolution of the effective gravitational
				constant. The values of the parameters used are similar to the Figure~(\ref{fig1}). 
			}
			\label{fig3}
		\end{center}
	\end{figure} 
\section{Perturbation in MOG cosmology}\label{sec:2}
A fundamental issue in cosmology is the formation of cosmic structures so, in order to efficiently distinguish between MOG and $\Lambda$CDM models, the
evolution of matter density perturbations must be compared with cosmological observations.\\
In the following given that in Section~(\ref{sec:1}) we considered the FRLW metric for determining dynamics of universe at the background level and also supposed that G is only dynamical scale field therefore we will consider the evaluation of perturbations in the linear regime ($\delta\ll 1$). However, we remind  other fields ($\mu$ and $\omega$) must be considered to study the nonlinear growth rate of structures($\delta> 1$).
 Since, we will survey the linear perturbations of the metric tensor and the matter and scalar field perturbations.\\
 The Jeans equation in the context of GR gives the density perturbation equation for the non-relativistic single-component fluid  as follow:
	\begin{eqnarray}\label{delta}
		\ddot{\delta}+2H\dot{\delta}+\left(\frac{c_s^2 a_0^2k^2}{a^2}-4\pi G(a)\rho\right)\delta=0, \;\label{per0}
	\end{eqnarray}
where $c_s=\sqrt{dp/d\rho}$ is the speed of sound and the quantity $k/a$ is called the comoving wave number and we assumed G(a) in the context of MOG~\citep{Moffat:2007ju}.\\
The factor sign in parentheses determines the nature of the solutions of Equation~\eqref{per0}. When the pressure is larger than the gravity term (in other words for large k) we have oscillatory solution and on the other hand when gravity term predominates (for small k) the density contrast is growing. 
When gravity and the pressure terms are equal, the Jeans wave number is given by:
	\begin{equation}
		k_J=\frac{a}{a_0}\frac{\sqrt{4\pi G(a) \rho}}{c_s};
	\end{equation}
corresponding to the Jeans length $\lambda_J = 2\pi/k_J$ (wavelength) and for non-relativistic matter $k_J\gg (a/a_0)H$ and $c_s\ll 1$.\\
$V_G$s do not couple to photons since they are electrically neutral  and they can be considered  as almost pressure-less,
so the pressure gradient term in the Jeans equation is absent
and $c_s \sim 0$. The wavelength is approximately zero and we obtain
	\begin{eqnarray}
		\ddot{\delta}+2H\dot{\delta}-4\pi G(a)\rho\delta=0. \;\label{per1}
	\end{eqnarray}
It is more appropriate to write equation~\eqref{per1} in terms of the scale factor as:
	\begin{equation}
		\delta^{''}(a)+ \left (\frac{3}{a}+\frac{H'(a)}{H(a)} \right)\delta^{'}(a)-\frac{3}{2}\frac{ \Omega_{b0} G(a)}{a^5 H^2(a)/H_0^2}\delta(a)=0\label{ddm1},
	\end{equation}
which is solved numerically taking an initial contrast of $\delta_{mi} = 5\times10^{-5}$ at an initial scale factor of $a_i = 10^{-4}$. This means that we study the evolution of structure in a matter dominated era and considering $G_{\rm eff}\approx 6G_N$.\\
In order to compare these models with observational data in perturbation level, we calculate the so-called linear growth rate, namely the logarithmic derivative of the linear density contrast with respect to the scale factor:
	\begin{equation}
		f(a)=\frac{d\ln \delta_m}{d\ln a} \,.
	\end{equation}
	\begin{figure}
		\begin{center}
			\includegraphics[width=8cm]{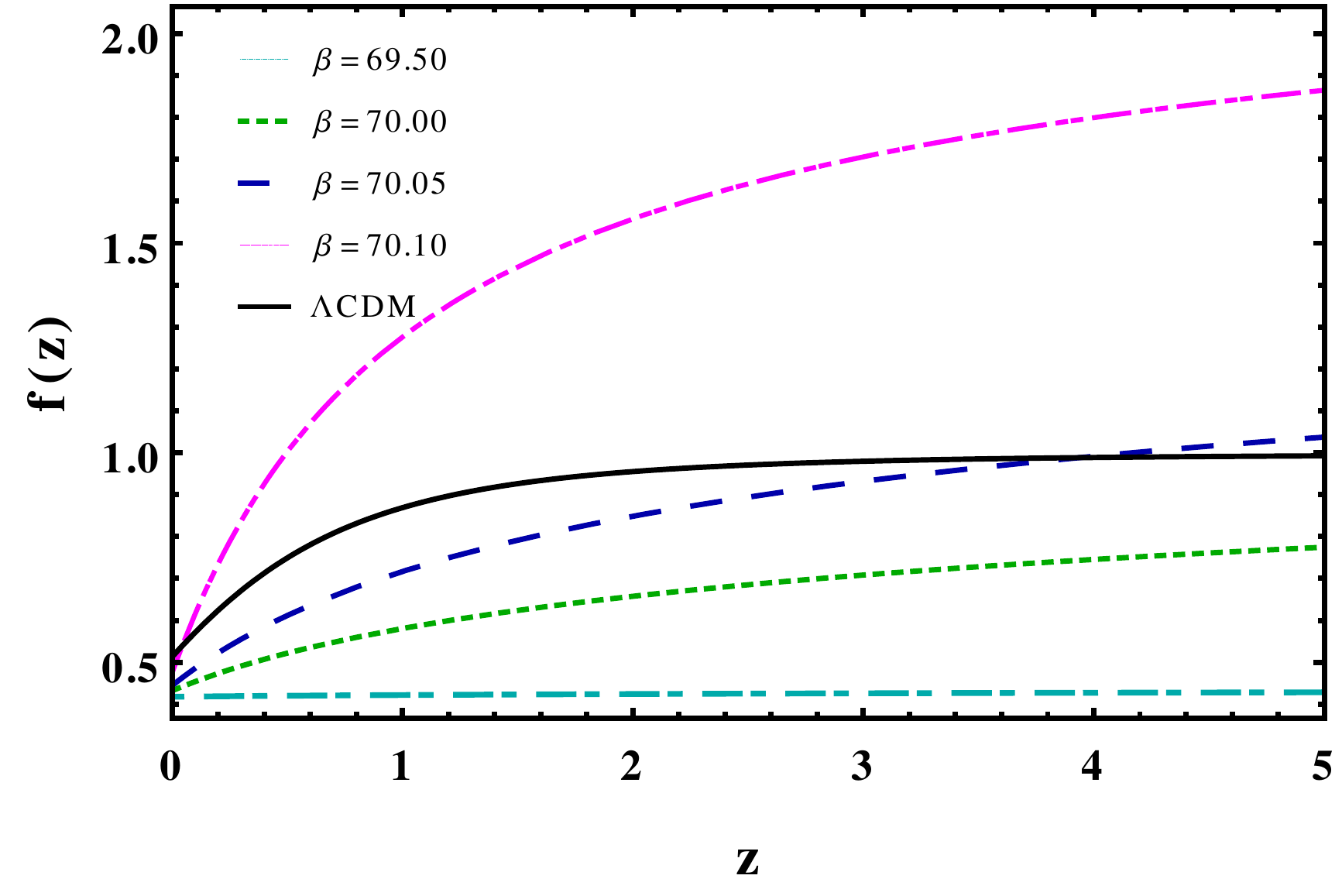}
			\includegraphics[width=8cm]{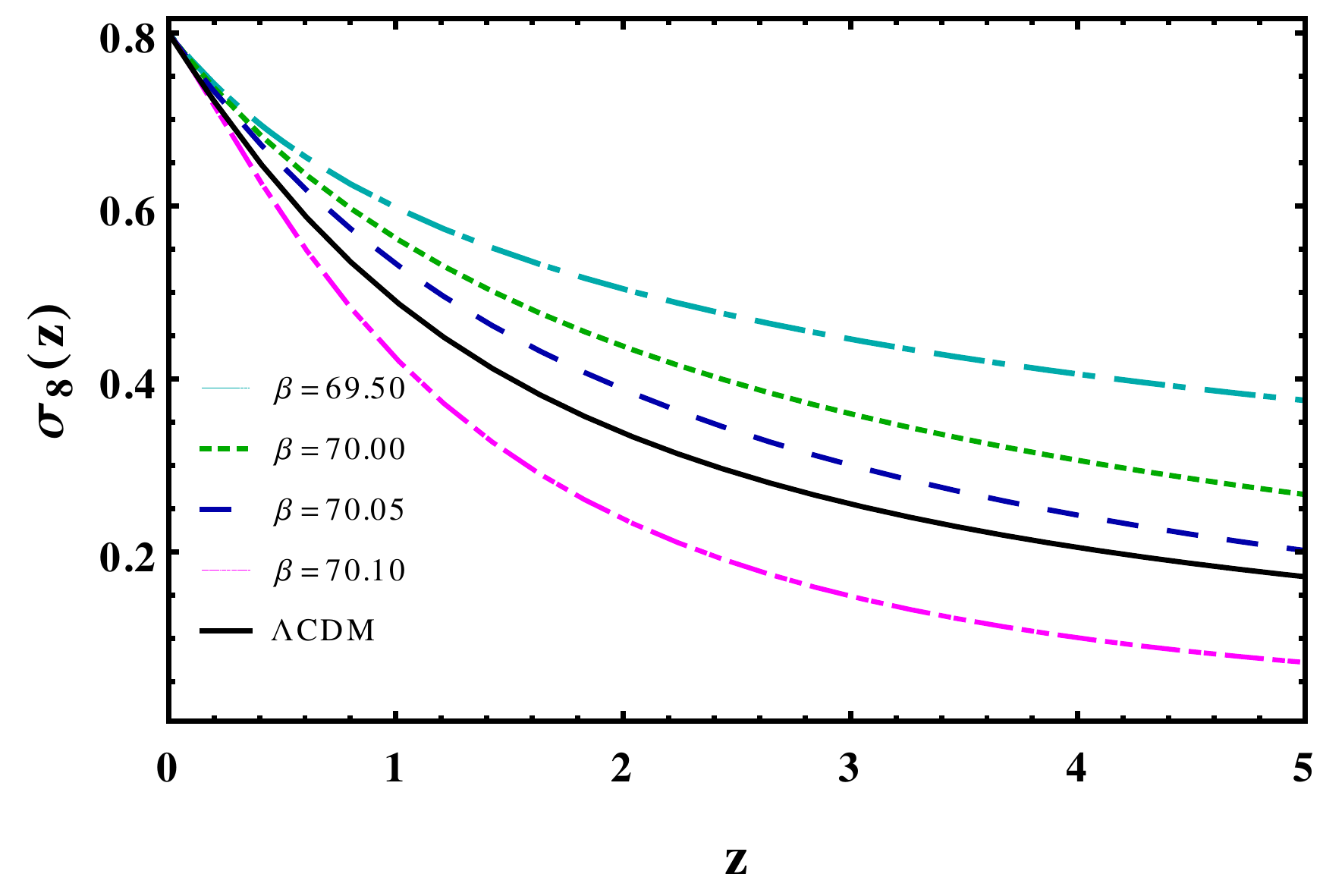}
			\includegraphics[width=8cm]{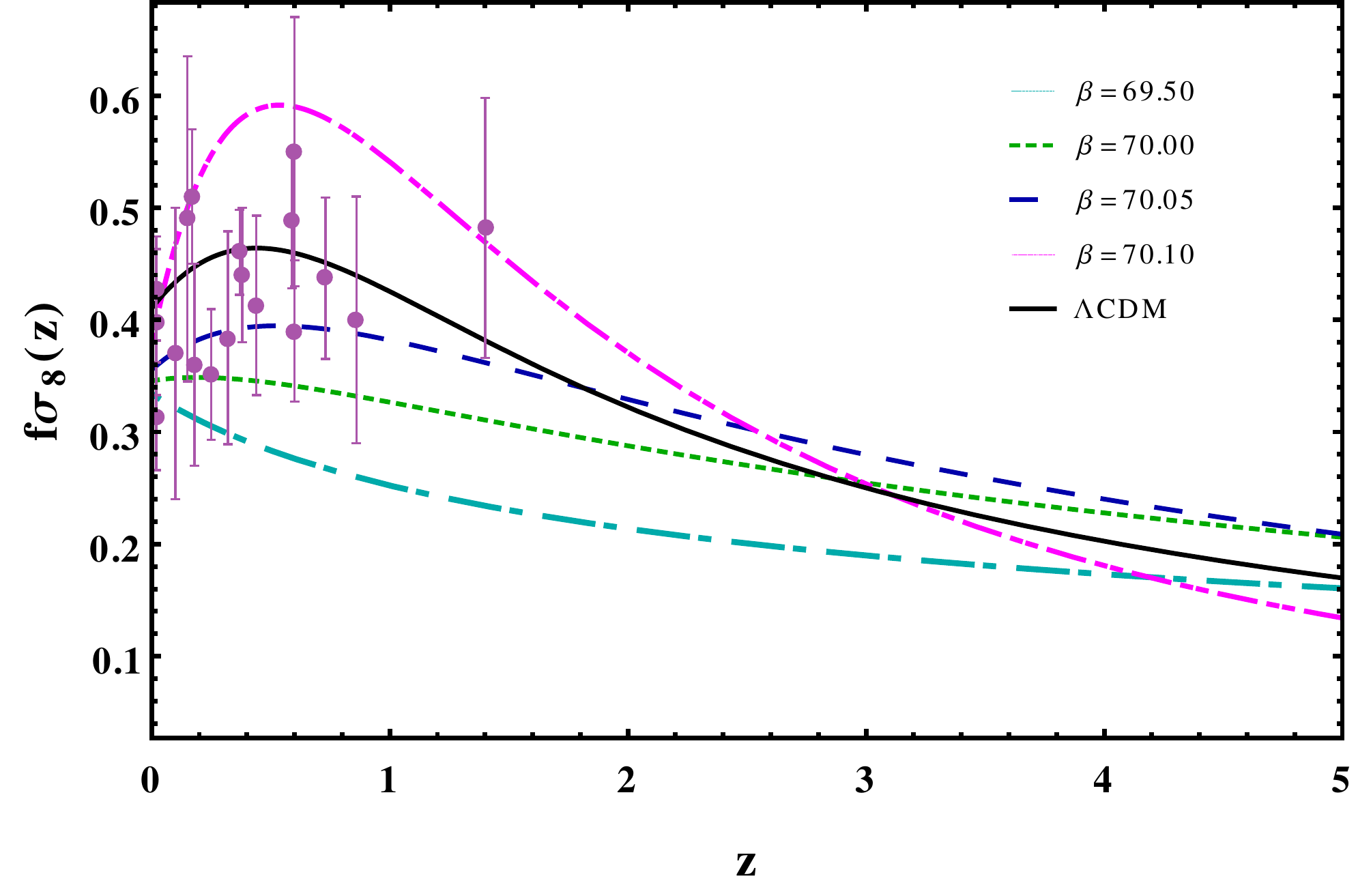}
			\caption{Redshift evolution of matter growth rate function $f(z)$ (top panel), variance of perturbations $\sigma_8(z)$ (middle panel) in the context and (bottom panel) the long term evolution of $f(z)\sigma_8(z)$ compared to observational growth rate data points in MOG cosmology\citep{Nesseris:2017vor}.  Here, we show the concordance $\Lambda$CDM model by solid black curve.
			}
			\label{fig4}
		\end{center}
	\end{figure} 
The growth rate has been observed in the redshift ranging $of z\in[0.02,1.4]$ by several surveys. In the top panel of Figure~(\ref{fig4}), we plot the linear growth factor $f(z)$ for the different values of $\beta$. We see in this panel that the amplitude of matter perturbations at low redshifts decreases due to the accelerated expansion of recent times in the universe. This panel shows that the growth of perturbations is negligible at high redshifts and consequently, the growth function goes to unity value, which corresponds to the matter dominated universe but it did not happen for value $\beta>70.08$. We observe that the suppression of the amplitude of matter fluctuations in the MOG cosmologies starts sooner as compared with the concordance $\Lambda$CDM model. Another important quantity in the perturbation level to consider is the redshift-dependent rms fluctuations of the linear density field within spheres of radius $R = 8h^{-1}$Mpc,  $\sigma_8(z) $ and it is computed as follows
\begin{equation}
\sigma_8(a)=\sigma_{8}(a=1)\frac{\delta_m(a)}{\delta_m(a=1)}.
\end{equation} 
We see the evolution of $\sigma_8(z)$ as a function of redshift $z$ in the middle panel of Figure~(\ref{fig4}). It shows that opposite to the behavior of growth rate function, $\sigma_8(z)$ in MOG models for $\beta<70.08$ is larger than the one in the case of the $\Lambda$CDM cosmology. A robust measurable quantity in perturbation surveys is $f(z)\sigma_8(z)=f\sigma_8(z)$. This quantity is shown in the bottom panel of Figure~(\ref{fig4}). As expected, the quantity $f(z)\sigma_8(z)$ is sensitive to the value of the parameters $\beta$. For simplicity, we will denote $\sigma_8(z=0)=\sigma_{80}$ when there is no ambiguity. It should be noted that in all panels of Figure~(\ref{fig4}) we adopted in addition the initial value of the  parameters similar to Figure~(\ref{fig1}), $\sigma_{80}=0.80$ for all models.\\

\section{MOG cosmology versus data}\label{sec:3}

In what follows, our aim is to constrain the parameters of the MOG and $\Lambda$CDM models using the currently available cosmological data at the background and perturbation levels via the MCMC method. We use the maximum likelihood estimation as the fitting method. In this method according to ${\cal L}_{\rm tot} \simeq e^{-\chi^2_{\rm tot}/2}$, the total likelihood function  is maximized by minimizing  $\chi^2_{tot}$ . To determine $\chi^2_{\rm tot}$ we use the following observational data set: big bang nucleosynthesis (BBN, 1 point), the Hubble evolution data (H(z), 31 points), type Ia supernova data (SNIa, 40 points from Pantheon binned sample), the cosmic microwave background radiation data (CMB, 2 points) and the baryon acoustic oscillation data (BAO, 14 points). Therefore $\chi^2_{\rm tot}$ for the first analysis is given by the relation
	\begin{equation}\label{eq:like-tot_chi}
		\chi^2_{\rm tot,1}({\bf p})=\chi^2_{\rm BBN}({\bf p})+\chi^2_{OHD}({\bf p})+\chi^2_{\rm SN}({\bf p})+\chi^2_{\rm CMB}({\bf p})+\chi^2_{\rm BAO}({\bf p})\;,
	\end{equation}
where the statistical vector ${\bf p}$ in this analysis is $(\Omega_{b0}, \Omega_{dm0},H_0,r_{\rm d,fid})$, $(\Omega_{b0},H_0,r_{\rm d,fid},V_{\rm G0},\beta)$ for $\Lambda$CDM, MOG models, respectively. We also include perturbation observables in our second analysis, specifically we use 18 independent data of the growth rate that were obtained from RSDs of various galaxy surveys so $\chi^2_{\rm tot,2}({\bf p'})=\chi^2_{\rm tot,1}({\bf p})+\chi^2_{\rm fs}({\bf p'})$. In second analysis, the statistical vector ${\bf p'}$ also includes $\sigma_{80}=\sigma_8(z=0)$ in addition to the parameters mentioned above for ${\bf p}$~\citep{Davari:2019tni}. We gauge the statistical strength of RSD data to comparing the result of two analyses. In the Table~(\ref{tabprior}), we listed the prior ranges of free parameters. 
In following we discuss each $\chi^2$ in detail.
	\begin{table}
		\centering
		\caption{Summary of the flat priors on the cosmological parameters assumed in this work.}
		\begin{tabular}{|c | c |}
			\hline 
			Parameter&Prior range\\
			\hline\hline
			$\Omega_{b0}$&$[0.01,0.2]$\\
			$\Omega_{dm0}$&$[0.01,0.5]$\\
            $H_0$&$[50,100]$\\
            $\sigma_{80}$&$[0.5,1]$\\
            $r_{\rm d,fid}$&$>0$\\
            $\beta$&$>69.5$\\
            $V_{G0}$&$>0$\\
			\hline 
		\end{tabular}\label{tabprior}
	\end{table}
\subsection{Big Bang Nucleosynthesis}
One of the most common used primordial elements for	constraining the baryon density is the deuterium abundance and the radiative capture of protons in deuterium to produce helium-3~\citep{Nunes:2020uex}. The experimental value for the reaction rate is calculated in~\cite{Serra:2009yp}, constraining the baryon density to $\Omega_{b0}h^2=0.022\pm0.002$. We adopt this value as the Gaussian prior likelihood in our analysis. The $\chi^2_{\rm BBN}$ is given by
\begin{equation}
\chi^2_{\rm BBN}({\bf p})=\frac{[\Omega_{b0}h^2-0.022]^2}{0.002^2}.
\end{equation}

\subsection{Type Ia Supernovae}

The observations of type Ia supernovae (SNIa) is an important probe to probe the expansion history of the universe. This observations cover the scales up to $z\leq2$. Using the the light-curve templates of SNIa, we can obtain the absolute magnitude of SNIa from the light curves. Currently, there are different collected compilations of SNIa as the Union 2.1 sample by~\cite{Suzuki:2011hu} consisting of 580 SNIa data within the range of $[0.015,1.41]$, and full  Joint Light-curve Analysis (JLA)  sample which contains 740 SNIa within the range of  $[0.01 ,1.2]$~\cite{Betoule:2014frx}.
 The largest combined sample of SNIa  which consist of a total of 1048 SNIa in  the redshift range $[0.01,2.3]$ presented by \cite{Scolnic:2017caz}, is called the Pantheon Sample. 
In this analyze, we use the binned Pantheon SNIa dataset, which contains 40 data points in range $0.014< z <1.61$. We marginalize over the nuisance parameter analytically in the distance estimates $M$. In  \cite{Camarena:2018nbr}, $\chi^2_{\rm SN}$ was introduced as: 
	\begin{equation}
	\chi^2_{\rm SN}({\bf p})=S_2({\bf p})-\frac{S_1^2({\bf p})}{S_0}+\ln\frac{S_0}{2\pi}+\ln |2\pi S_{SN}|,
	\end{equation}
here
	\begin{eqnarray}
	&&S_0=V.S_{SN}^{-1}.V^T,\nonumber\\ 
	&&S_1=W.S_{SN}^{-1}.V^T,\nonumber\\ 
	&&S_2=W.S_{SN}^{-1}.W^T,\nonumber
	\end{eqnarray}
where covariance matrix, S is from the binned Pantheon sample (considering both systematic and statistical errors) and V is a row vector of unitary elements and $W_i=\mu_{\rm b,i}-\mu(z_i)$. The distance modulus, $\mu$, whose theoretical prediction is obtained in term the luminosity distance $d_L$ via:
	 \begin{equation}
	 \mu(z)=5\log_{10}\frac{d_L(z)}{10\rm pc}.
	 \end{equation}
In this way for this set of data, the cosmology-independent normalization constants can be dropped.\\
Table~(\ref{tabsn}) represent the values of the two common parameters obtained with the help of the sum of SNIa and BBN data. As we observe from Table (\ref{tabsn}), the SNIa data allow a very large value of the Hubble constant
at present and it is actually very close to its local measurements($H_0 = 74.03 \pm 1.42$).
	\begin{table}
		\centering
		\caption{The best values of the two common parameters obtained with a combination of the binned Pantheon SNIa dataset (40 points) and BBN data.}
		\begin{tabular}{|c | c c c|}
			\hline 
			Model&$\Omega_{b0}$&$H_0$&$\chi^2$\\
			\hline
			MOG&$0.044^{+0.001}_{-0.001}$&$70.34^{+1.6}_{-1.6}$&$67.75$\\
			\hline
			$\Lambda$CDM &$0.040^{+0.003}_{-0.006}$ &$74.35^{+6.5}_{-6.5}$&$67.24$	\\
			\hline 
		\end{tabular}\label{tabsn}
		\end{table}
\subsection{Hubble Observational Data}
The Observational Hubble Data (OHD) is a conventional test to check cosmological models which gives a direct measurement of the history of the expansion of the universe. The OHD sample is obtained from the differential age of galaxies technique (DAG), and measurements of peaks of BAO. 
The H(z) is determined with this method  from the following equation:
	\begin{equation}
	H(z)=\frac{-1}{1+z}\frac{dz}{dt}.
	\end{equation}
It is worth noting that data points derived from the cosmic chronometers approach, i.e. the massive and passively evolving galaxies ~\citep{Jimenez:2001gg} can be used as DAG technique which is cosmological-model-independent. 
This method is based on the differential age evolution of old elliptical passive-evolving galaxies that are separated by a small redshift interval but formed at the same time. These galaxies have not had any star formation ever since as they have been formed  in the early universe, at high redshift, with large stellar mass >$10^{11}\cal{M}_{\odot}$. In this way, by calculating the age difference of these galaxies, the derivative $dz/dt$ can be measured from the ratio $\Delta z/\Delta t$. $\Delta z$ obtained with high accuracy and straightforward from spectroscopic  observations.  The determination of $\Delta t$ is much more challenging, and this required standard and readable clocks that exist all over the universe so this was the main idea of a cosmic chronometer that stellar evolution could provide such standard clocks. 
If we found an extremely massive, old, and passively evolving stellar population across a range of redshifts since its stars had formed all simultaneously and evolved passively therefore that would be appropriate to consider as the cosmic chronometer~\citep{Moresco:2018xdr}.\\
 In this analysis, we use the 31 data points from the  recent accurate estimates of $H(z)$ in the redshift range $0.07\leq z \leq 1.965$. These data are uncorrelated with the BAO data points and was presented in \cite{Marra:2017pst}.\\
In this case we can write:
	\begin{equation}
		\chi^2_{OHD} ({\bf p})=\sum_{i}\frac{\left[ H(z_i,{\bf p})-H_i \right]^2}{\sigma_i^2} \,.
	\end{equation}
where $\sigma_i$ is the Gaussian error on the measured value of $H_i$. In the Tables~(\ref{tabh}), we reported the results of our statistical analysis for two models by using OHD and BBN data. As we see two models are consistent with observational data.
\begin{table}
		\centering
		\caption{The best values of the two common parameters obtained by using  OHD(31 points) and BBN data.}	
		\begin{tabular}{|c | c c c|}
			\hline 
			Model&$\Omega_{b0}$&$H_0$&$\chi^2$\\
			\hline
			MOG&$0.048^{+0.001}_{-0.001}$&$68.06^{+0.83}_{-1.6}$&$15.52$\\
			\hline
			$\Lambda$CDM &$0.048^{+0.01}_{-0.01}$ &$68.15^{+2.9}_{-2.1}$&$14.50$	\\
			\hline 
		\end{tabular}\label{tabh}
\end{table}
\subsection{Cosmic Microwave Background} 
The CMB is one of the most significant observables in cosmology. The advantage of CMB data is that 
it has well-understood linear physics  and precision to determine the cosmological parameters. Here, we will consider the compressed CMB likelihood (Planck 2018 TT, TE, EE + lowE) from \citet[][Table I]{Chen:2018dbv} on the angular scale of the sound horizon at the last scattering $l_a$ and the baryon density $\Omega_b h^2$. We do not use the spectral index since it does not appear directly in our calculation and also do not use the shift parameter, R since it depended on dark matter density. $l_a$ is given by:
	\begin{equation}
		l_a=(1+z_\star) \pi \frac{D_A(z_\star)}{r_s(z_\star)},
	\end{equation}
where $r_s(z)$ is  the comoving sound horizon distance at the recombination is given by  $r_s(z) =\int_z^\infty c_s(z)/H(z)$ that the sound speed, $c_s(z)=[3(1+3\frac{\Omega_{b0}}{4\Omega_{\gamma0}(1+z)})]^{-1/2}$.
We set $\Omega_{\gamma0}=2.469\times 10^{-5}h^{-2}$~\citep{Hinshaw:2012aka}.
The recombination redshift $z_\star$ is obtained using the fitting function proposed by~\cite{Hu:1995en} as:
	\begin{equation}\label{zcmb}
		z_\star=1048[1+0.00127(\Omega_{b0}h^2)^{-0.738}][1+g_1(\Omega_{dm0}h^2)^{g_2}],
	\end{equation}
where $g_1=(0.0783(\Omega_{b0}h^2)^{-0.238})/(1+39.5(\Omega_{b0}h^2)^{0.763})$ and $g_2=(0560)/(1+21.1(\Omega_{b0}h^2)^{1.81})$. It should be noted that for the MOG model, since there is no dark matter, only the first term in the equation~(\ref{zcmb}) is considered. In~\cite{Aghanim:2018eyx} reported its value $z_\star=1089.80\pm0.21$ on Planck $\rm TT,\rm TE,\rm EE+\rm lowE+\rm lensing$. Then we can define $\chi^2_{CMB}$ as
	 	\begin{equation}
	 	\chi^2_{\rm CMB}({\bf p})=X^T_{\rm CMB}. C_{\rm CMB}^{-1}. X_{\rm CMB}.
	 	\end{equation}
For MOG model, we used a prior on $(l_a, \Omega_{b_0}h^2)$  for the $w$CDM model derived from the Planck 2018 results~\citep{Chen:2018dbv}:
	\[X_{\rm CMB}= \left(
	\begin{array}{l l}
	l_a - 301.462\\
	\Omega_{b_0}h^2 - 0.02239\\
	\end{array} \right), \]
	and
	\begin{equation}\label{ccmbmog}
		C_{\rm CMB}^{-1} = 
		\left({\begin{array}{cc} 139.594&21357.9\\21357.9&28267800\end{array}}\right).
	\end{equation}
	and also these value for $\Lambda CDM$ be considered as
	\[X_{\rm CMB}= \left(
	\begin{array}{l l}
	l_a - 301.471\\
	\Omega_{b_0}h^2 - 0.02236\\
	\end{array} \right), \]
	and
	\begin{equation}\label{ccmbland}
		C_{\rm CMB}^{-1} = 
		\left({\begin{array}{cc} 141.675&27740\\27740&49875900\end{array}}\right).
	\end{equation}
We show the values of the two common parameters for two models obtained by using CMB and BBN data in the Table~(\ref{tabcmb}). A noteworthy point in these results is the large difference between the values of $H_0$ obtained for the two models.
	\begin{table}
		\centering
		\caption{The best values of the two common parameters obtained by using CMB and BBN data.}	
		\begin{tabular}{|c | c c c|}
			\hline 
			Model&$\Omega_{b0}$&$H_0$&$\chi^2$\\
			\hline
			MOG&$0.048^{+0.001}_{-0.001}$&$68.51^{+0.42}_{-0.52}$&$0.04$\\
			\hline
			$\Lambda$CDM &$0.044^{+0.001}_{-0.002}$ &$70.01^{+1.8}_{-1.5}$&$0.03$	\\
			\hline 
		\end{tabular}\label{tabcmb}
	\end{table}
\begin{table*}
	\centering
	\caption{The table shows the distance constraints of BAO measurements for various observational probes. The redshift, the mean value, standard deviation, and the corresponding reference of the observable in each case are reported in the table, respectively.}	
	\begin{tabular}{|c | c c c|}
		\hline 
		$z_{\rm eff}$&$\rm Value$&$\rm Observable$&$\rm Reference$\\
		\hline\hline
		$0.15$&$(664\pm25.0)(r_d/r_{\rm d,fid})Mpc$&$D_V$&\cite{Ross:2014qpa}\\
		$0.44$&$(1716\pm83.0)(r_d/r_{\rm d,fid})Mpc$&$D_V$&\cite{Kazin:2014qga}\\
		$0.6$&$(2221\pm101.0)(r_d/r_{\rm d,fid})Mpc$&$D_V$&\cite{Kazin:2014qga}\\
		$1.52$&$(3843\pm147.0)(r_d/r_{\rm d,fid})Mpc$&$D_V$&\cite{Ata:2017dya}\\
		\hline
		$0.81$&$(1649.5\pm66.0)(r_d/r_{\rm d,fid})Mpc$&$D_A$&\cite{Abbott:2017wcz}\\
		\hline
		$0.38$&$(1512\pm33.0)(r_d/r_{\rm d,fid})Mpc$&$D_M$&\cite{Alam:2020sor}\\ 
		$0.51$&$(1975\pm71.0)(r_d/r_{\rm d,fid})Mpc$&$D_M$&\cite{Alam:2020sor}\\
		$0.61$&$(2307\pm80.0)(r_d/r_{\rm d,fid})Mpc$&$D_M$&\cite{Alam:2020sor}\\
		$2.3$&$(5566\pm317.2)(r_d/r_{\rm d,fid})Mpc$&$D_M$&\cite{Bautista:2017zgn}\\
		$2.4$&$(5259.7\pm250.5)(r_d/r_{\rm d,fid})Mpc$&$D_M$&\cite{Riess:2018uxu}\\
		\hline
		$2.3$&$(1336\pm45.7)(r_d/r_{\rm d,fid})Mpc$&$D_H$&\cite{Bautista:2017zgn}\\
		$2.4$&$(1327.4\pm53.0)(r_d/r_{\rm d,fid})Mpc$&$D_H$&\cite{Riess:2018uxu}\\
		\hline
		$0.38$&$81.2\pm3.2\rm km s^{-1}\rm Mpc^{-1}$&$H$&\cite{Alam:2020sor}\\
		$0.51$&$90.9\pm3.3\rm km s^{-1}\rm Mpc^{-1}$&$H$&\cite{Alam:2020sor}\\
		\hline
	\end{tabular}\label{tabbaodata}
\end{table*}
\subsection{Baryon acoustic oscillations}
The equilibrium between the pressure and gravity in the baryonic matter of cosmic fluid results in 
periodic fluctuations of the density which is so called the Baryonic Acoustic Oscillation (BAO).  
BAO provides a standard ruler to measure cosmological distances.
 The BAO signal directly constrains the Hubble parameter H(z) along the line of sight and also BAO constrains the angular diameter distance $D_A(z)=\frac{1}{1+z}\int_0^zdz'/H(z')$ in a redshift shell. 
 The Dark Energy Survey in first data release provides a measurement of $D_A$
  at $z_{\rm eff} = 0.81$, using the projected two point correlation function of a sample of over 1.3 million galaxies with
 measured photometric redshifts, distributed over a footprint of 1336 deg$^2$~\citep{Abbott:2017wcz}.\\
 BAO is determined in the 2D correlation function while $H(z)$ is determined in 3D space. In some case it is impossible to combine both quantities can be measured as
  \begin{equation}
 D_V(z)=\left[(1+z)^2D^2_{\rm A}(z)\frac{z}{H(z)}\right]^{\frac{1}{3}}.
 \end{equation}
 We present in the Table~(\ref{tabbaodata}), the isotropic angular diameter distance $D_V$ for some $z_{\rm eff}$ from surveys \citep{Ross:2014qpa, Kazin:2014qga}.
 
 The final galaxy clustering data release of the Baryon Oscillation Spectroscopic Survey~\citep{Alam:2020sor}, provides measurements of the comoving angular diameter distance $D_M=(1 + z)D_A(z)$ and Hubble parameter H at effective redshifts 0.38, 0.51, and 0.61 from the BAO method after applying a reconstruction method. Another parameter that we have used is the Hubble distance, $D_H(z)=1/H(z)$ from the Sloan Digital Sky Survey
 (SDSS) data release 12 (DR12)~\citep{Bautista:2017zgn} and the final
 data release of the SDSS-III~\citep{Riess:2018uxu}.
 
 In order to measure the BAO scale from the clustering of matter, the definition of fiducial cosmology is required so approximately the distance constraints presented in Table~(\ref{tabbaodata}) are multiplied by a factor $(r_d /r_{\rm d,fid} )$, which is the ratio between the sound horizon to the same quantity computed in the fiducial cosmology. 
 We take this ratio as a free parameter in the statistical analysis for both models. We use 14 uncorrelated data points of the Table~\ref{tabbaodata} in our analysis.
In Table~(\ref{tabbao}), we reported the results of our statistical analysis for two models by using BAO data.
	\begin{table}
		\centering
		\caption{The best values of the two common parameters obtained by using BAO and BBN data.}	
		\begin{tabular}{|c | c c c|}
			\hline 
			Model&$\Omega_{b0}$&$H_0$&$\chi^2$\\
			\hline
			MOG&$0.047^{+0.001}_{-0.002}$&$66.50^{+0.52}_{-0.91}$&$17.35$\\
			\hline
			$\Lambda$CDM &$0.047^{+0.002}_{-0.003}$ &$68.28^{+1.8}_{-1.8}$&$13.08$	\\
			\hline 
		\end{tabular}\label{tabbao}
	\end{table}	
\subsection{Redshift Space Distortions }\label{sec-rsd}
 One of the very important probes of large scale structure is redshift space distortions that they provide measurements of $f\sigma_8(a)$. In galaxy redshift surveys, RSD measures the peculiar velocities of matter. The result is inferring the growth rate of cosmological perturbations. This measurement is done on a wide range of redshifts and scales.
This can be obtained by measuring the ratio of the monopole and the quadrupole multipoles of the redshift space power spectrum which depends on $\beta = f /b$. Here $f$ is the growth rate and $b$ is the bias The combination of $f\sigma_8(a)$ is independent of the bias factor and the bias dependence in this combination cancels out thus. This combination is a good discriminator of the models~\citep{Nesseris:2017vor, universe5060137}.\\
In this study, we use the vigorous $f\sigma_{8} (z)$ measurements from the "Gold-2017" compilation given in \citet[][Table III]{Nesseris:2017vor}. 
	\begin{table}
		\centering
		\caption{The best values of the two common parameters obtained with a combination of the $f\sigma_{8}(z)$ dataset (18 points) and BBN data.}
		\begin{tabular}{|c | c c c c|}
			\hline 
			Model&$\Omega_{b0}$&$H_0$&$\sigma_{80}$&$\chi^2$\\
			\hline
			MOG&$0.047^{+0.005}_{-0.006}$&$68.61^{+0.8}_{-1.2}$&$0.79^{+0.06}_{-0.05}$&$11.95$\\
			\hline
			$\Lambda$CDM &$0.036^{+0.007}_{-0.005}$ &$78.12^{+6.4}_{-4.9}$&$0.81^{+0.05}_{-0.05}$&$11.53$	\\
			\hline 
		\end{tabular}\label{tabfs8}
		\end{table}
In this case we can write:
	\begin{equation}
	\chi^2_{fs} ({\bf p})=\sum_{i}\frac{\left[ f\sigma_{8}(z_i,{\bf p})-f\sigma_{8,i} \right]^2}{\sigma_i^2} \,.
	\end{equation}
In Table~(\ref{tabfs8}), we reported results of the best values of the two common parameters obtained with a combination of the $f\sigma_8 (z)$ dataset and BBN data.
\begin{table*}
	\centering
	\caption{Marginalized constraints (1$\sigma$ uncertainties) of free parameters for two models using the background observables of equation \eqref{eq:like-tot_chi}.}
	\begin{tabular}{|c | c|  c|c| c|  c|c|}
		\hline 
		Model&$\Omega_{b0}$&$\Omega_{m0}$&$H_0$& $V_{G0}$&$\beta$&$r_{d,fid}$\\
		\hline\hline
		MOG&$ 0.049^{+0.002}_{-0.002}$&$-$&$ 67.80^{+1.0}_{-1.1}$&  $0.89^{+0.02}_{-0.08}$&$70.06^{+0.002}_{-0.002}$&$182.4^{+2.9}_{-2.1}$\\
		\hline
		$\Lambda$CDM&$0.044^{+0.001}_{-0.001}$ &$0.26^{+0.01}_{-0.02}$& $71.10^{+1.0}_{-1.0}$&$-$& $-$&$149.3^{+2.5}_{-2.5}$	\\
		\hline 
	\end{tabular}\label{tab:res:back}
\end{table*}
\begin{table}
	\centering
	\caption{Model selection of the statistical results by using  the background observables of the equation~\eqref{eq:like-tot_chi}.}
	\begin{tabular}{|c | c| c|  c|c|c|c|}
		\hline
		Model&$\chi^2_{\rm min}$&$\chi^2_{\rm red}$ &$\Delta \chi^2_{\rm min}$  &$\Delta$AIC&$\Delta$BIC&$\Delta$DIC\\
		\hline \hline
		MOG&$110.1$&$1.33$&$3.7$& $5.9$& $8.2$&$3.4$\\
		\hline
		$\Lambda$CDM&$106.4$&$1.27$&$-$&$-$&$-$\\
		\hline 
	\end{tabular}\label{chi-back}
\end{table}	
	\begin{figure*}
		\begin{center}
			\includegraphics[width=12cm]{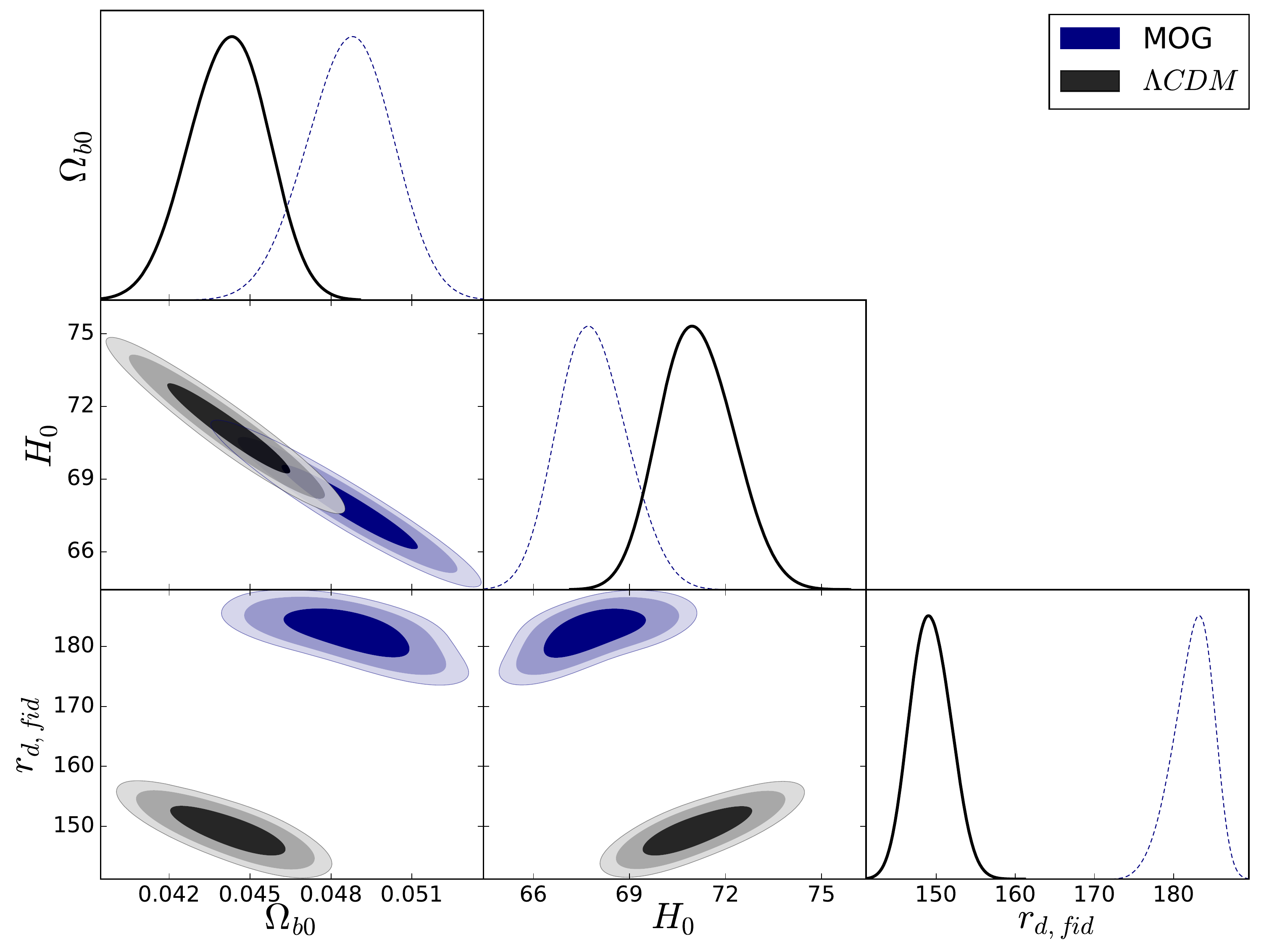}
			\caption{2D contours and 1D posterior distributions of the free
				parameters of MOG model(blue contour) that are in common with the $\Lambda$CDM model(black contour) relative to background data. using  background data(see equation \eqref{eq:like-tot_chi}). See Table~(\ref{tab:res:back}) for the numerical values.
			} 
			\label{fig-con-back}
		\end{center}
	\end{figure*} 
	\begin{figure*}
		\begin{center}
			\includegraphics[width=12cm]{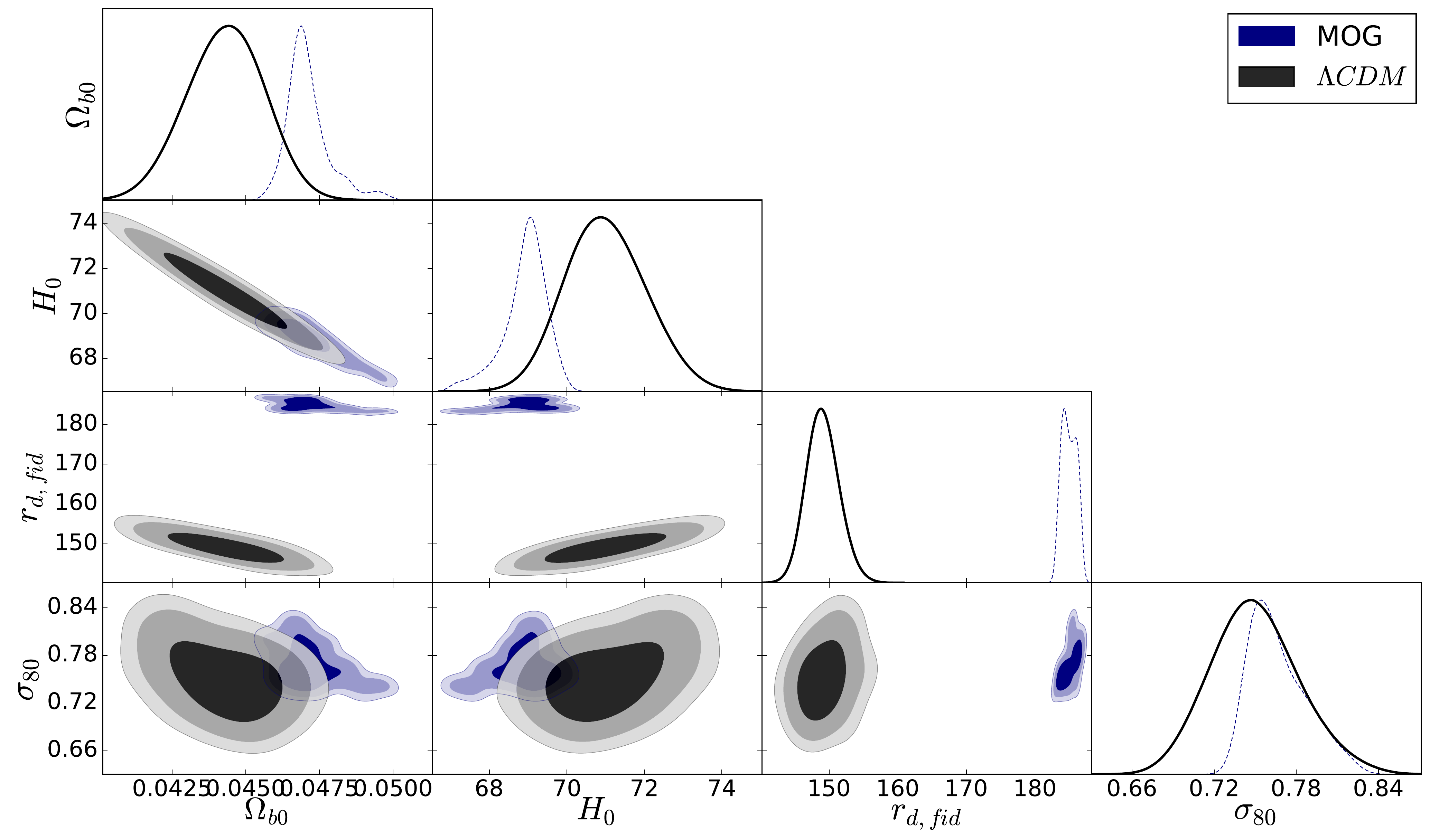}
			\caption{ 2D contours and 1D posterior distributions of the free
				parameters for the MOG model(top left panel) and $\Lambda$CDM model using both background and perturbation data.
			} 
			\label{fig-con-tot}
		\end{center}
	\end{figure*} 	
\subsection{Joint analysis}	
\begin{figure}	
		\begin{center}
			\includegraphics[width=8cm]{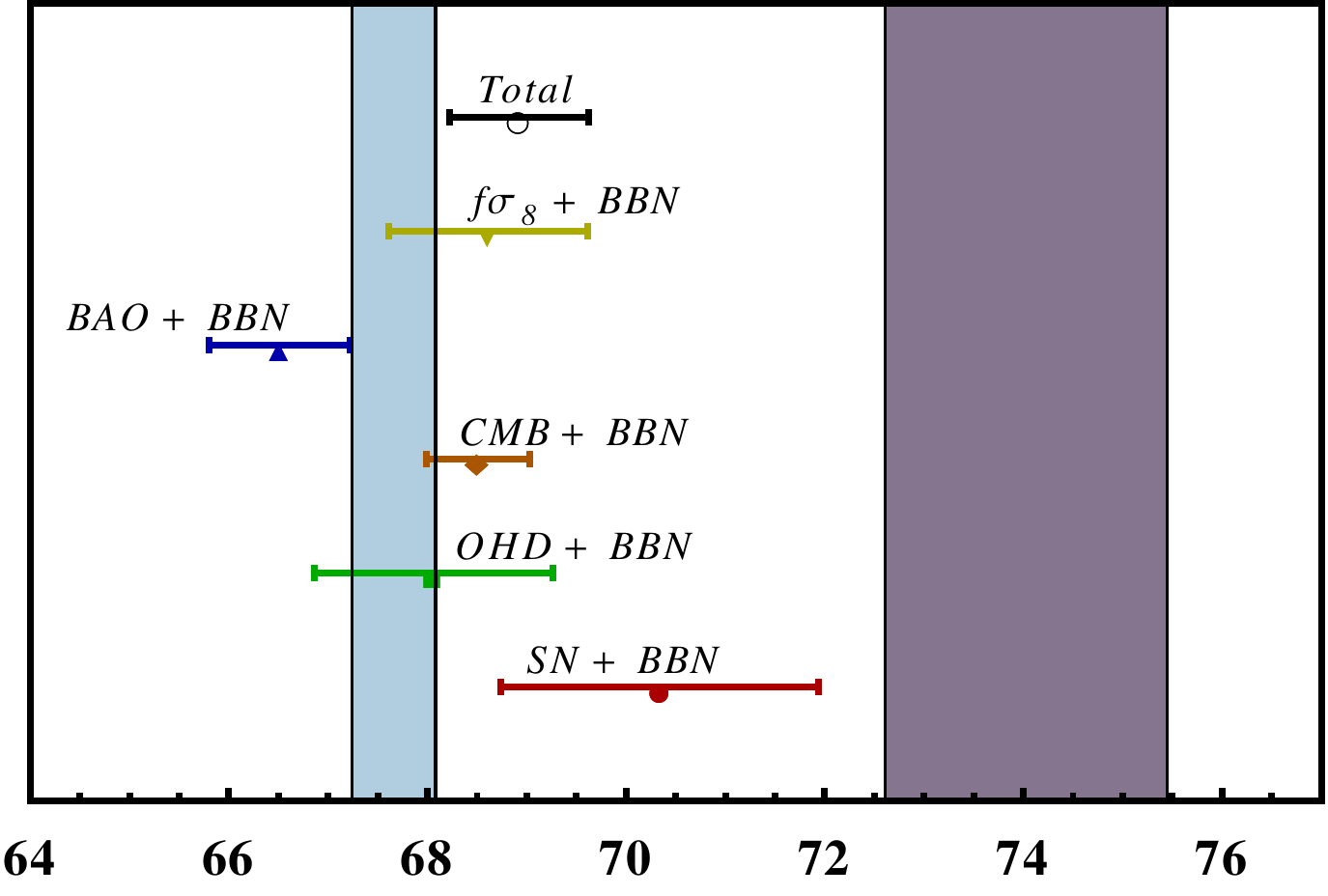}
			\includegraphics[width=8cm]{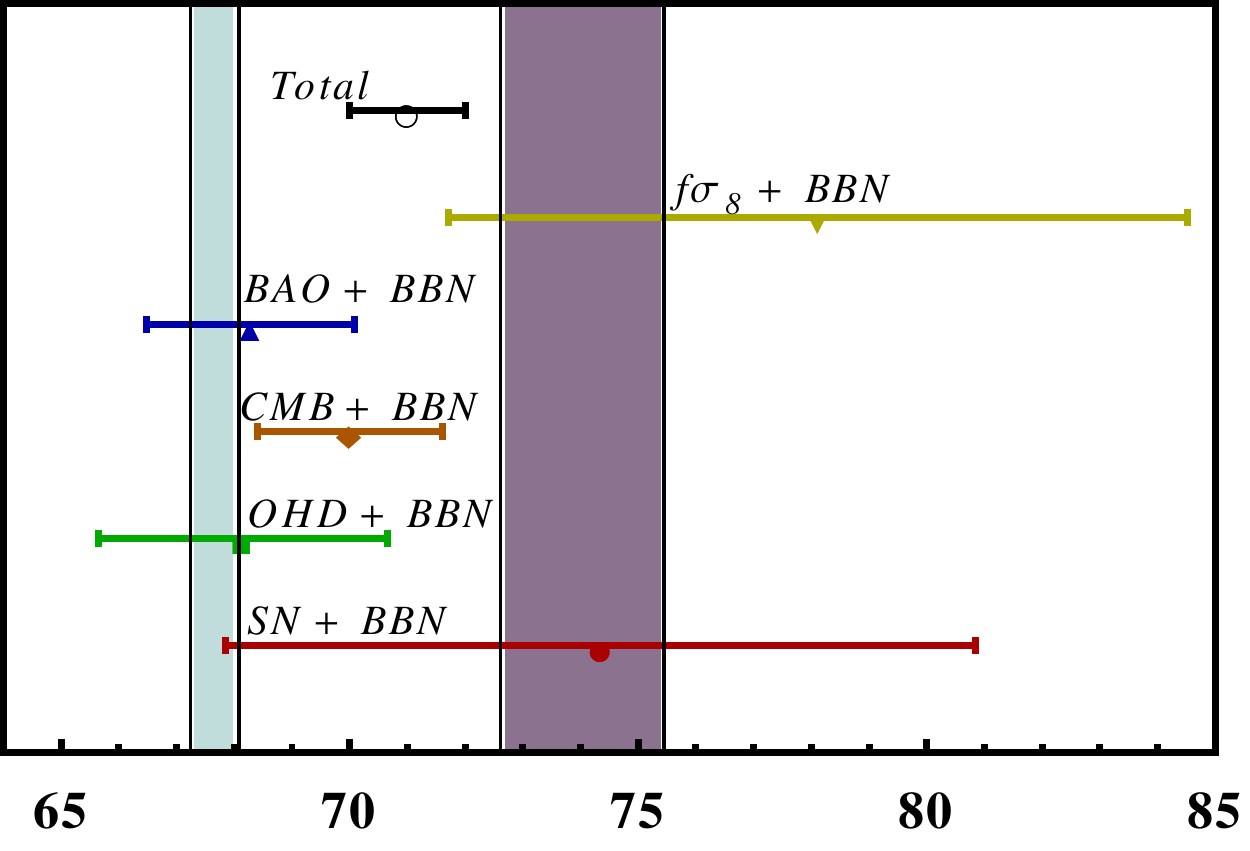}						\caption{Whisker graph with the 68\% confidence level region for $H_0$ for the observational datasets considered in this work as separately and combination for MOG (top panel) and $\Lambda$CDM (bottom panel) models. 
				Schematic representation of $H_0$ at 1$\sigma$ ( SN (red color), OHD(green color), CMB(orange color), BAO(blue color), $f\sigma_{8}$(yellow color) and Total: SN+OHD+BAO+CMB+BBN+$f\sigma_{8}$(black  color)) for MOG (top panel) and $\Lambda$CDM (bottom panel) models. 
				We also show
				Planck 2018, $H_0 = 67.66 \pm 0.42 \textrm{ km s}^{-1} \textrm{Mpc}^{-1}$(cyan region) by
				\citet{Aghanim:2018eyx} and $H_0 = 74.03 \pm 1.42 \textrm{ km s}^{-1} \textrm{Mpc}^{-1}$(gray region) by \citet{Riess:2019cxk}.
			}\label{figh0}
		\end{center}
	\end{figure}

In this work, we wish to compare the MOG model against the novel observational data. Specifically, we perform an overall statistical analysis using the background(geometrical) data (SN, BAO, CMB, Big Bang Nucleosynthesis, OHD data) and the growth data. So in the first joint analyses, we considered the background data and performed a likelihood analysis by the MCMC algorithm. We listed the best fit values of parameters using the joint analysis of geometry measurements data for MOG and $\Lambda$CDM models in Table~(\ref{tab:res:back}). As can be seen, there is a discrepancy between the best values of the two models for $H_0$. It seems MOG theory predicted a smaller value for $H_0$ which is in disagreement with the local observations of $H_0$ like HOLiCOW ($H_0$ Lenses in COSMOGRAIL's Wellspring) experiment in~\cite{Wong:2019kwg}. They have used a joint analysis of six gravitationally lensed quasars with measured time delays to achieve the highest-precision probe of $H_0$ and found $H_0=73.3\pm1.8$ for a flat $\Lambda$CDM cosmology. Recently \cite{Birrer:2020tax} based on a new hierarchical Bayesian analyses in which the mass sheet transform is only constrained by stellar kinematics have reported new results. They used mock lenses, which are generated from hydrodynamic simulations and in first step, they applied the inference to the TDCOSMO sample of seven lenses, six of which are from H0LiCOW, and measured $H_0=74.5^{+5.6}_{-6.1}$. In the second step, they added spectroscopy and imaging for a set of 33 strong gravitational lenses from the Sloan Lens ACS (SLACS) sample in order to more constrain the deflector mass density profiles and they resolved kinematics to constrain the stellar anisotropy for nine of the 33 SLAC lenses. Assuming that the TDCOSMO and SLACS galaxies are derived from the same parent population they measured $H_0=67.4^{+4.1}_{-3.2}$ from the joint hierarchical analysis of the TDCOSMO+SLACS sample so we see that the result of the MOG theory is consistent with their reported results. Of course, this difference in the values reported of \cite{Birrer:2020tax} may be due to that the H0LiCOW could be hiding a correlation with the CMB dipole direction as is disccused  in~\cite{Krishnan:2021dyb}. They presented the cordinates of lenses on the celestial sphere from H0LiCOW and declination and their value of $H_0$.\\
 In the Figure~(\ref{fig-con-back}) we show the $1\sigma$ and $2\sigma$
confidence levels for the parameters $\Omega_{b0}$, $H_0$ and $r_{\rm d,fid}$ that are in common with the $\Lambda$CDM model.
As we see, the contours for $\Omega_{b0}$ and $H_0$ overlap, and the contour of $\Omega_{b0}$ for the MOG model has expanded to larger values and inversely $H_0$ for $\Lambda$CDM wides to higher values. Interestingly, there is a significant difference between $r_{\rm d,fid}$ contours for the two models.\\
In the second analysis, we also include perturbation observables, 18 RSD independent data points. We repeated statistical analyses and the observational constraints are summarized in Table~(\ref{tab:res:tot}). Contour of the common parameters of the two models in this case, $(\Omega_{b0},H_0,r_{\rm d,fid},\sigma_{80})$, are shown in the Figure~(\ref{fig-con-tot}). Interestingly, the contours in the MOG case shift towards a higher value of $\Omega_{b0}$ and smaller values of $H_0$. We also see that $\sigma_{80}$ for the MOG model relative to $\Lambda$CDM shift towards higher value, It is in agreement with other works as~\cite{galaxies1010065}, \cite{SHOJAI201743}, where MOG increases the growth rate of matter perturbations, compared to $\Lambda$CDM.\\ 
Additionally, in Figure~(\ref{figh0}) we concisely display the constraints on the present value of the Hubble parameter$H_0$, considering the observational datasets considered in this work as separately and combination.
\section{Model selection}\label{sec:4}
Model selection is the process of choosing and distinguishing candidate models with various free parameters. To select the most compatible model with the observational data, we need numerical procedures that can obtain the goodness of fit and also determine the best value of free parameters. The least squares method is one of the simplest comparisons which have mostly been applied in cosmology.
In general, the $\chi^2$ method is  very popular and sufficient for comparing different models when they have the same number of degree of freedom(d.o.f) ,and the model with a smaller $\chi^2$ means it has a better fit with the data.
However, for models with a different d.o.f, a model with more parameters tends to lead to the lowest $\chi^2$. Under these conditions, the $\chi^2$ method is unrighteous for comparing models. So in order to test the quality of fit of the present models,
an important quantity that is used for the data fitting process is the reduced $\chi^2$ as
\begin{align}
\chi^2_{\rm red} &= \frac{\chi^2_{\rm min}}{N-K} \,, \label{red}
\end{align}
where $N$  denotes the total number of data points used in the statistical analysis and $K$ the number of free parameters for the model. 
As an approximate method, when the variance of the measurement error is known a priori, a $\chi _{\rm red }^{2}\gg 1$ indicates a poor model fit which the fit has not entirely taken the data.
Having the value of $\chi _{\rm red }^{2}$ around $1$  indicates that the extent of the match between observations and estimates is in accord with the error variance. A $\chi^2_{\rm red}\ll 1$ indicates that the model is "over-fitting" the data: either the model is inadequately fitting noise, (or the error variance has been overestimated)~\citep{2003drea}.

Therefore we conclude from the result of the Table~(\ref{chi-back}) the MOG theory ($\chi^2_{\rm red}=1.3$)  and $\Lambda$CDM model ($\chi^2_{\rm red}=1.2$) are consistent with the observational background data. But in some studies such as~\cite{Andrae:2010gh} was discussed that the traps involved in using $\chi^2_{\rm red}$. There are two independent problems: (a) the number of degrees of freedom can only be estimated for linear models and inverse for nonlinear models, The number of dof is unknown i.e., it is impossible to calculate the value of reduced  $\chi^2_{\rm red}$. (b) The value of $\chi^2_{\rm red}$ itself is subject to noise due to random noise in the data, i.e., its value is uncertain. Since this uncertainty impairs the efficiency of reduced chi-squared for differentiating between models or recognizing the convergence of a minimization procedure so we consider two kinds of criteria for model selection: the Akaike Information Criterion (AIC) and Bayesian Information Criterion (BIC) for comparing two models as:
\begin{align}
{\rm AIC}&=\chi^2_{\rm min}+2K+\frac{2K(K+1)}{N-K-1} \,, \label{aic}\\
{\rm BIC}&=\chi^2_{\rm min}+K\ln N \,. \label{bic}
\end{align}

Since $\Lambda$CDM has the best agreement to almost observational data
so $\Lambda$CDM is the best choice to consider a reference model. Now, for any model $M$, other than the reference model (denoted by
R), from the difference $\Delta \rm AIC = |\rm AIC_M - \rm AIC_R$|, we
get at the following conclusions of 
(i) If $\Delta \rm AIC\leq2$ , then the concerned model has substantial support with respect to the reference model (i.e.
it has evidence to be a good cosmological model), (ii)
$4\leq \Delta \rm AIC\leq7$ indicates less support with respect
to the reference model, and finally, (iii) $\Delta \rm AIC\geq 10$
means that the model has no support, in fact, it has no use in principle.
Similarly, for the BIC a difference in the range  $0<\Delta \rm BIC\leq2$  is considered as weak evidence, for  $2< \Delta \rm BIC\leq6$ is considered as positive evidence, and $6<\Delta \rm BIC\leq 10$  is strong evidence while more than 10 is very strong evidence against the model with the higher BIC.\\ 
Another additional information criterion that we describe in the following is the Deviance Information Criterion (DIC). \cite{SBCL02} developed this method of model selection in an effort to generalize the AIC.
The DIC has some interesting properties. First, it could use for the situation where one or more parameters or a combination of parameters are insignificantly constrained by the data such as in astrophysics, unlike the AIC and BIC.
Secondly, it is easily measured  from posterior samples, such as
those generated by MCMC methods. It has already been used to study quasar clustering in
astrophysics , and it is stated as~\citep{Liddle:2007fy,Rezaei:2021qpq}:
\begin{equation}
DIC=2\overline{\chi^2_{\rm tot}({\bf p})}-\chi^2_{\rm tot}(\bar{{\bf p}}).
\end{equation}
Therefore in the first analysis that we use the background observational data, the results of Table~(\ref{chi-back}) show that the MOG model has less support compared to the reference model, $\Lambda$CDM with observational background data ($\Delta \rm AIC$) and that there is positive evidence against it ($\Delta \rm BIC$). Then the evidence in favor of the MOG model becomes a category weaker and indicating strong support for the $\Lambda$CDM model. We can conclude from  $\Delta$DIC that both models are
equally supported by the data.\\
In the second analysis, For the background and the growth rate data, our results in Table~(\ref{chi-tot}) especially $\Delta$DIC show that MOG 
and $\Lambda$CDM models fit the cosmological data equally well.
	\begin{table*}
		\centering
		\caption{Marginalized constraints (1$\sigma$ uncertainties) for free parameters of two models using both the background and perturbation data.}
		\begin{tabular}{|c | c|  c|c| c|  c|c|c|}
			\hline 
			Model&$\Omega_{b0}$&$\Omega_{m0}$&$H_0$& $V_{G0}$&$\beta$&$r_{d,fid}$&$\sigma_{80}$\\
			\hline\hline
			MOG&$ 0.047^{+0.001}_{-0.001}$&$-$&$ 68.92^{+0.7}_{-0.4}$&  $0.80^{+0.02}_{-0.01}$&$70.06^{+0.001}_{-0.001}$&$185.0^{+1.1}_{-1.1}$&$0.77^{+0.02}_{-0.03}$\\
			\hline
			$\Lambda$CDM&$0.044^{+0.001}_{-0.001}$ &$0.26^{+0.01}_{-0.01}$& $71.0^{+1.0}_{-1.0}$&$-$& $-$&$149.2^{+2.2}_{-2.5}$&$0.75^{+0.03}_{-0.04}$	\\
			\hline 
		\end{tabular}\label{tab:res:tot}
	\end{table*}
\begin{table}
	\centering
	\caption{Model selection of the statistical results by using the background and perturbation observables.}
	\begin{tabular}{|c | c| c|  c|c|c|c|}
		\hline
		Model&$\chi^2_{\rm min}$&$\chi^2_{\rm red}$ &$\Delta \chi^2_{\rm min}$  &$\Delta$AIC&$\Delta$BIC&$\Delta$DIC\\
		\hline \hline
		MOG&$122.1$&$1.22$&$2.3$& $4.5$& $6.9$&$0.07$\\
		\hline
		$\Lambda$CDM&$119.8$&$1.19$&$-$&$-$&$-$&\\
		\hline 
	\end{tabular}\label{chi-tot}
\end{table}	
\section{Cosmological evolution}\label{sec:5}
\begin{figure}
		\begin{center}
			\includegraphics[width=8cm]{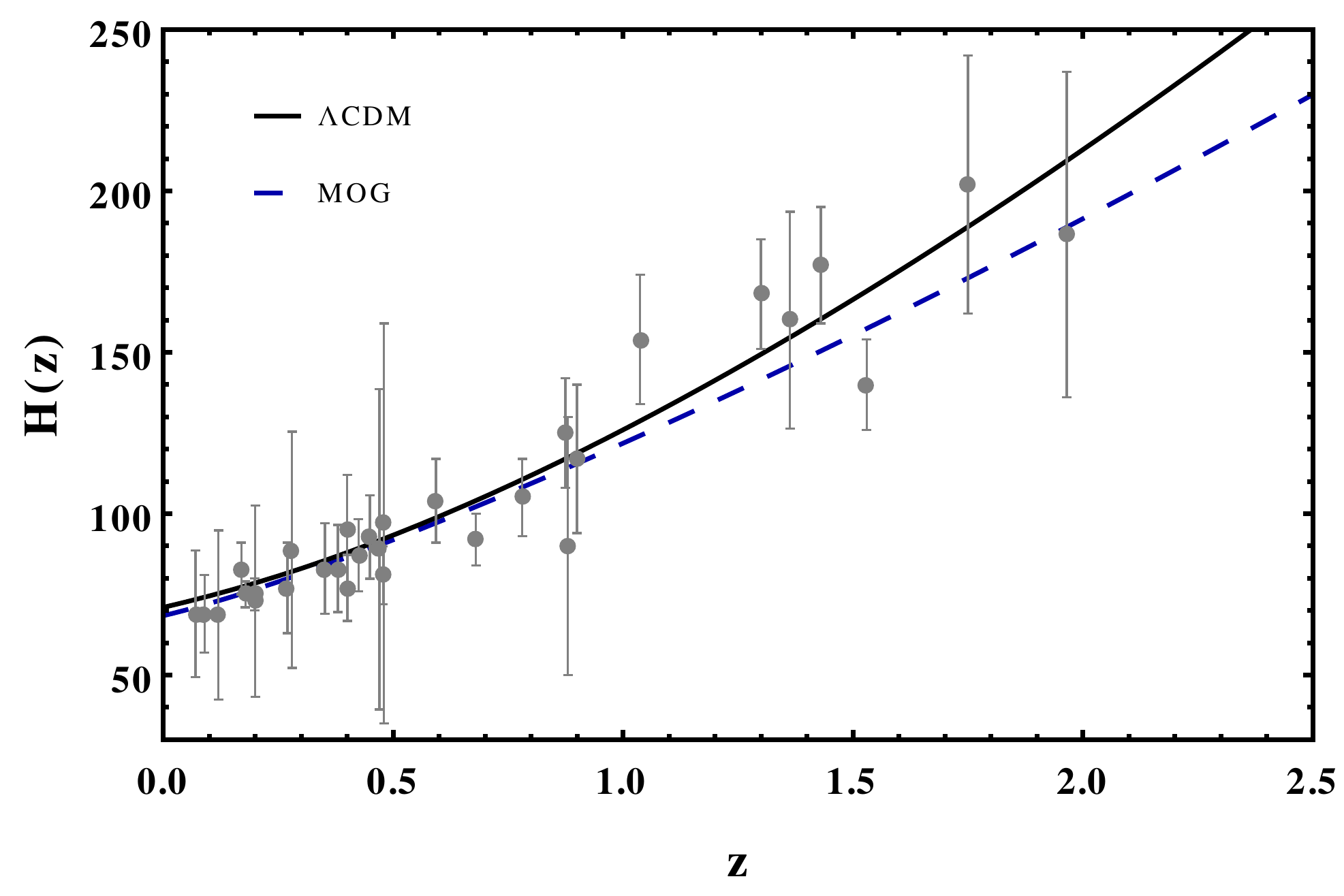}
			\includegraphics[width=8cm]{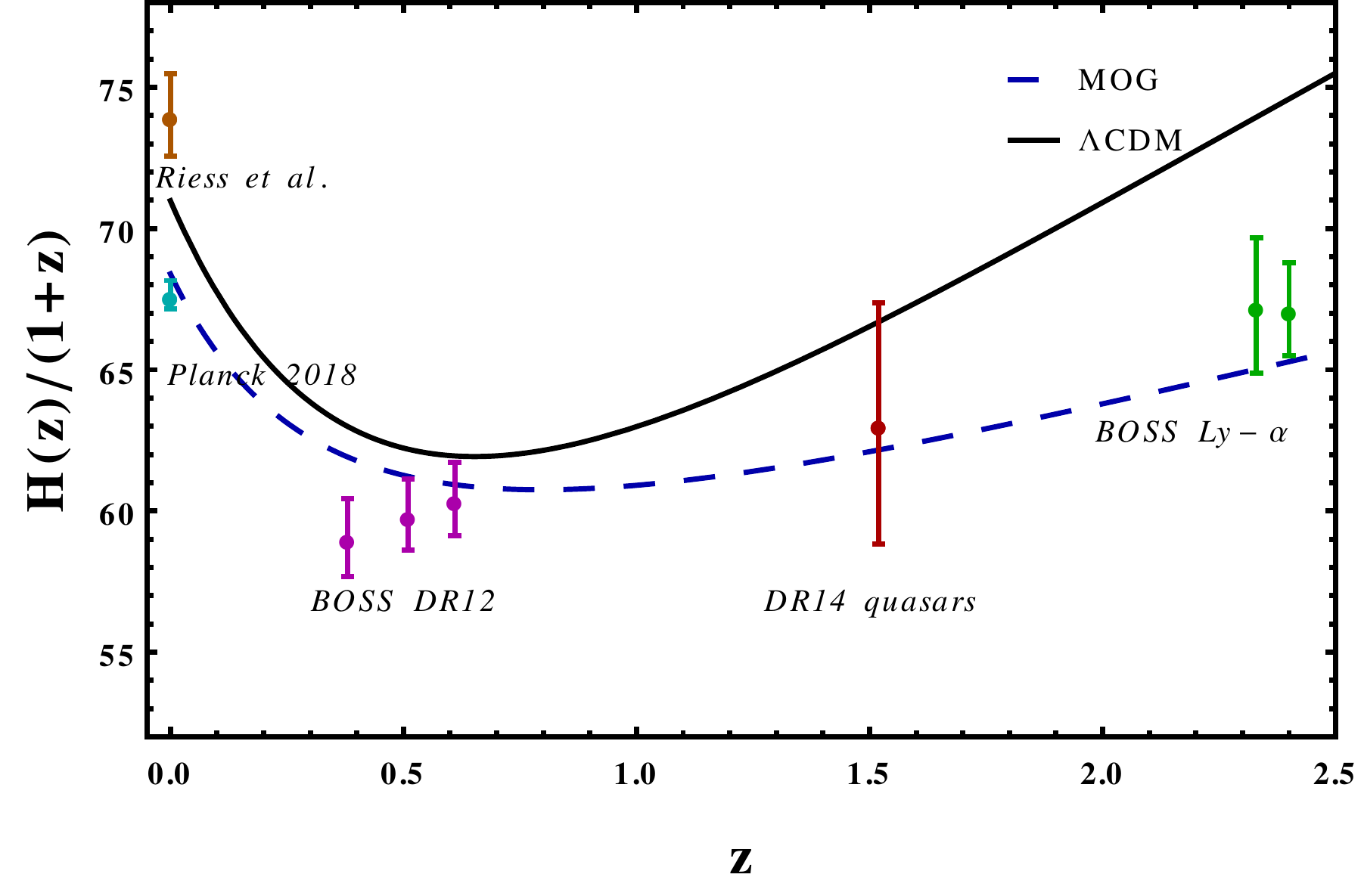}						\caption{In the top panel, we compared the 31 cosmic chronometer data points in gray dots~\citep{Marra:2017pst} and theoretical redshift evolution of Hubble parameter H(z). In the bottom panel, we have plotted H(z) vs. redshift. The black solid curve shows the concordance $\Lambda$CDM model.
			}	\label{fighbest1}
		\end{center}
	\end{figure}

\begin{figure}	
			\begin{center}
				\includegraphics[width=8cm]{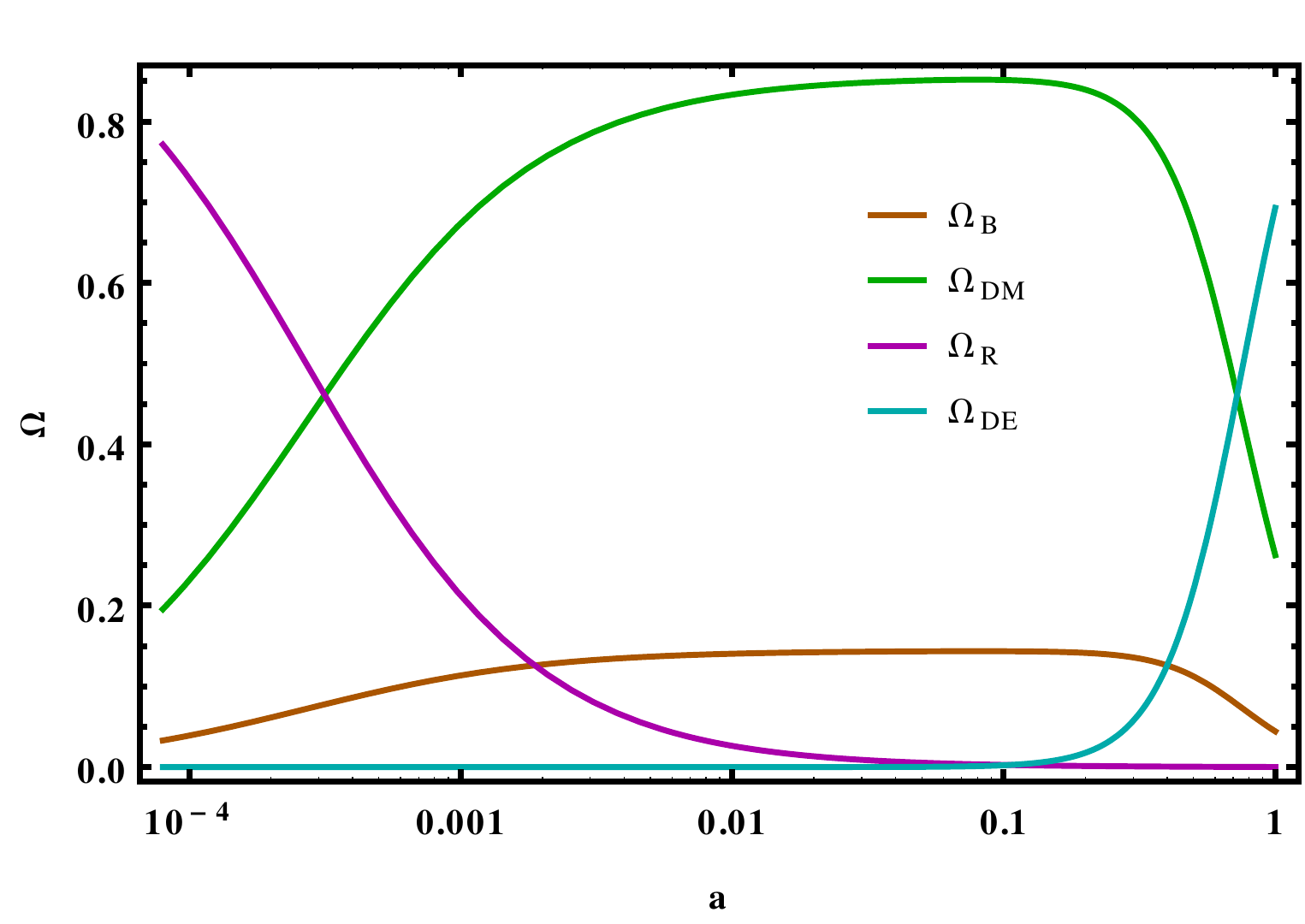}
				\includegraphics[width=8cm]{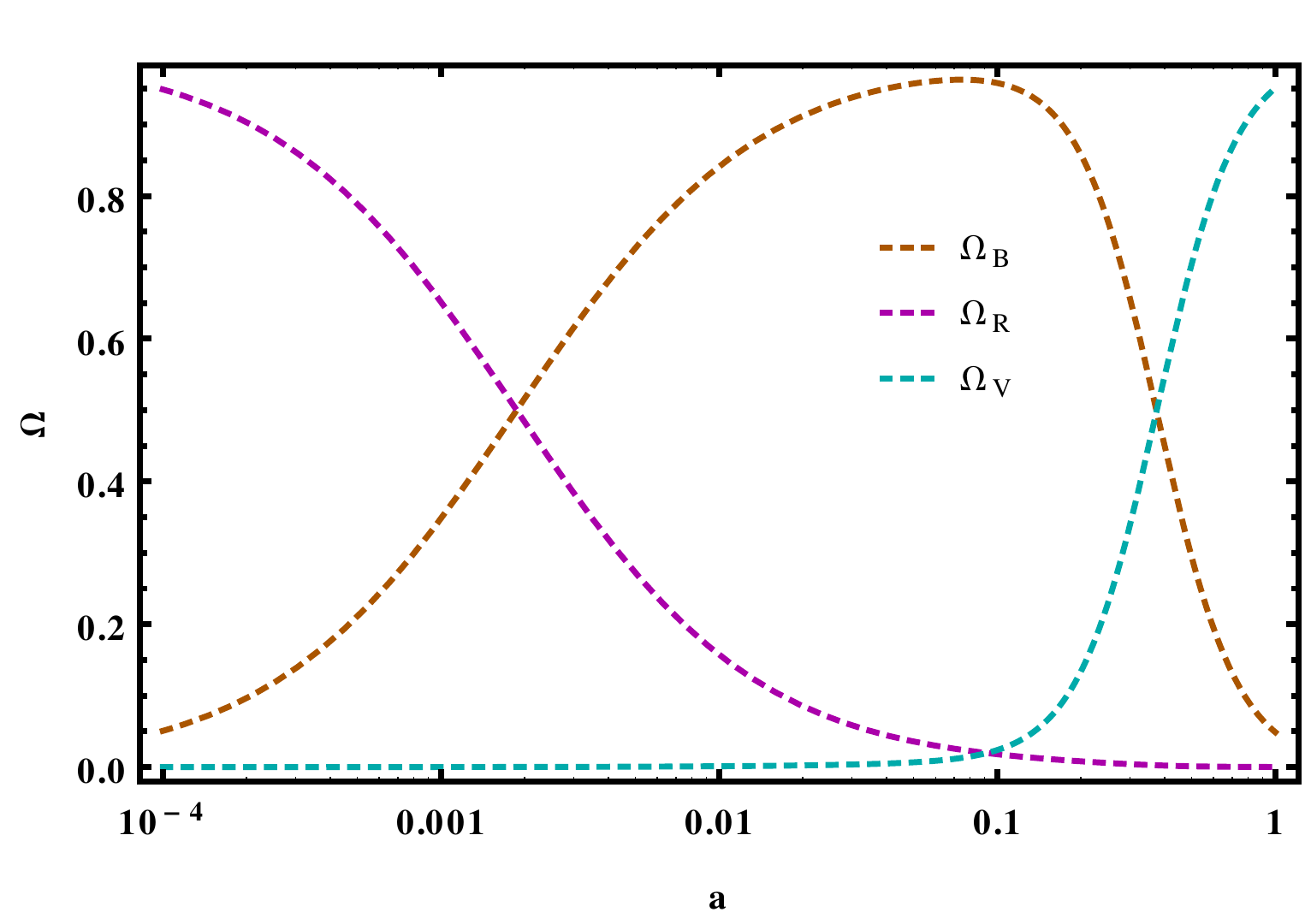}						\caption{Evolution of the fractional energy density of radiation (violet curves), baryonic matter (orange curves), and DE	component (blue curves) in terms of a cosmic scale factor for standard $\Lambda$CDM(top panel) and MOG(bottom panel) cosmologies. The solid green curve represents dark matter component of the concordance $\Lambda$CDM model. In both models, we use the best fit values from Table~(\ref{tab:res:tot}).
				}
				\label{figomegabest}
			\end{center}
		\end{figure}

In this section, we illustrate the evolution of basic cosmological values based on the best appropriate values of the cosmological parameters presented in Table~(\ref{tab:res:tot}).\\
We plot the evolution of H(z) and $H(z)/(1+z)$ in Figure~(\ref{fighbest1}) for best fit parameters in Table~(\ref{tab:res:tot}). \\
In Figure~(\ref{figomegabest}), we present the evolution of the fractional energy densities for different components. We see for two models, the universe evolves from a radiation dominated phase to a matter dominated epoch and the density parameter of the pressure-less matter decreases and dark energy for $\Lambda$CDM model and $V_G$ in the MOG model increases by decreasing redshift, indicating the early time matter dominated universe.
\begin{figure}
	\begin{center}
		\includegraphics[width=8cm]{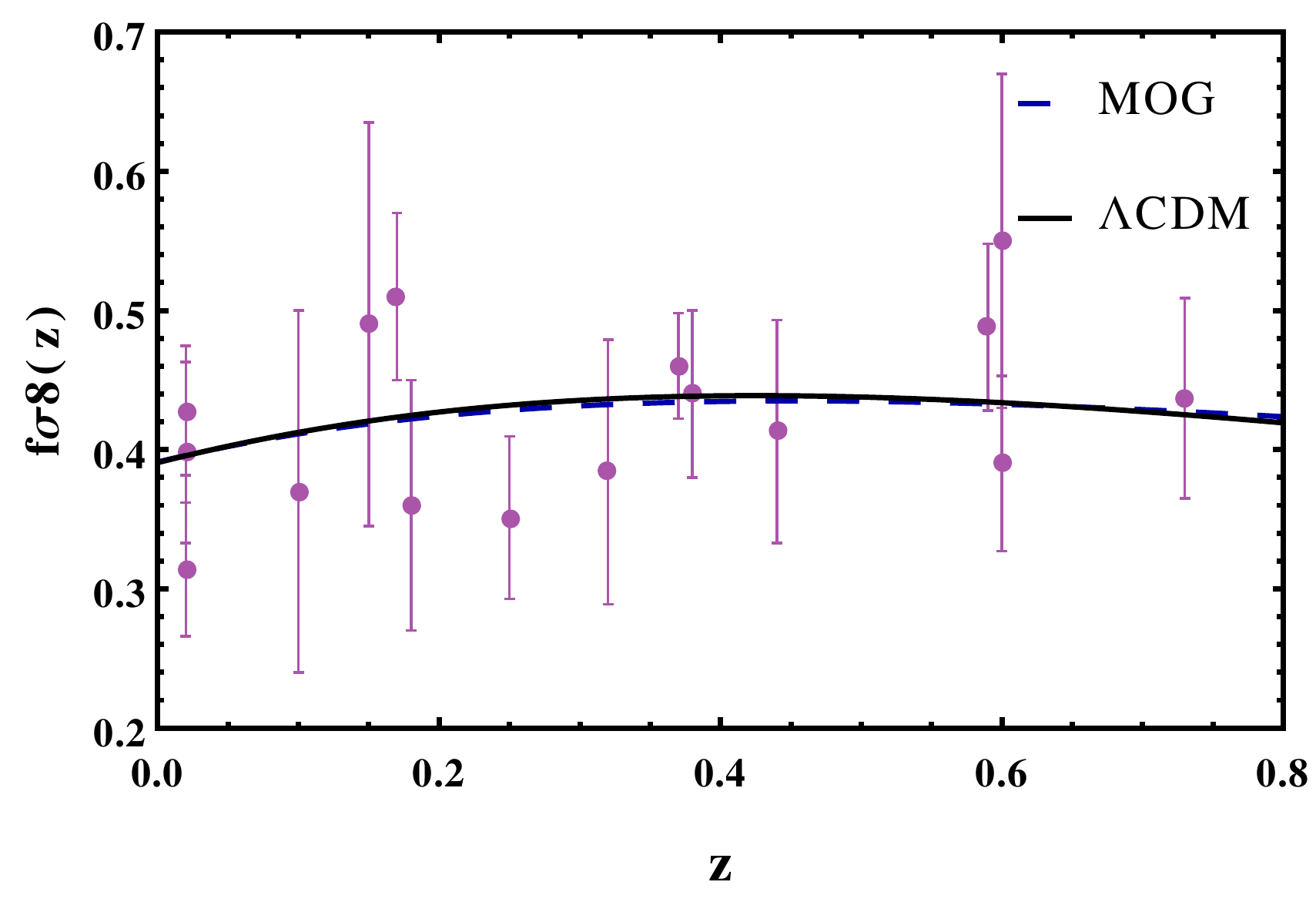}
			\caption{ The evolution of $f\sigma_ 8 (z)$ as a function of the cosmic redshift and the "Gold-2017" growth rate dataset. The blue dashed and the black solid curves correspond to the best fits of MOG and $\Lambda$CDM models, respectively. See Table~(\ref{tab:res:tot}) for the numerical value of free parameters.} 
		\label{fig-fs8best}
	\end{center}
\end{figure} 
In Figure~(\ref{fig-fs8best}), we compare the observed $f\sigma_{8}(z)$ with the theoretical value of the growth rate function for two models by the best free parameters.
As we expected from statistical analysis in section~(\ref{sec-rsd}) and the results of Table~(\ref{chi-tot}), the predicted growth rate in the MOG model can  fit the observed $f\sigma_{8}(z)$ as well as $\Lambda$CDM model. It is consistent with the results of \cite{galaxies1010065} where they modeled perturbation growth in MOG theory and compared the observed matter power spectrum with the prediction at the present time. 

\begin{figure}
	\begin{center}
		\includegraphics[width=8cm]{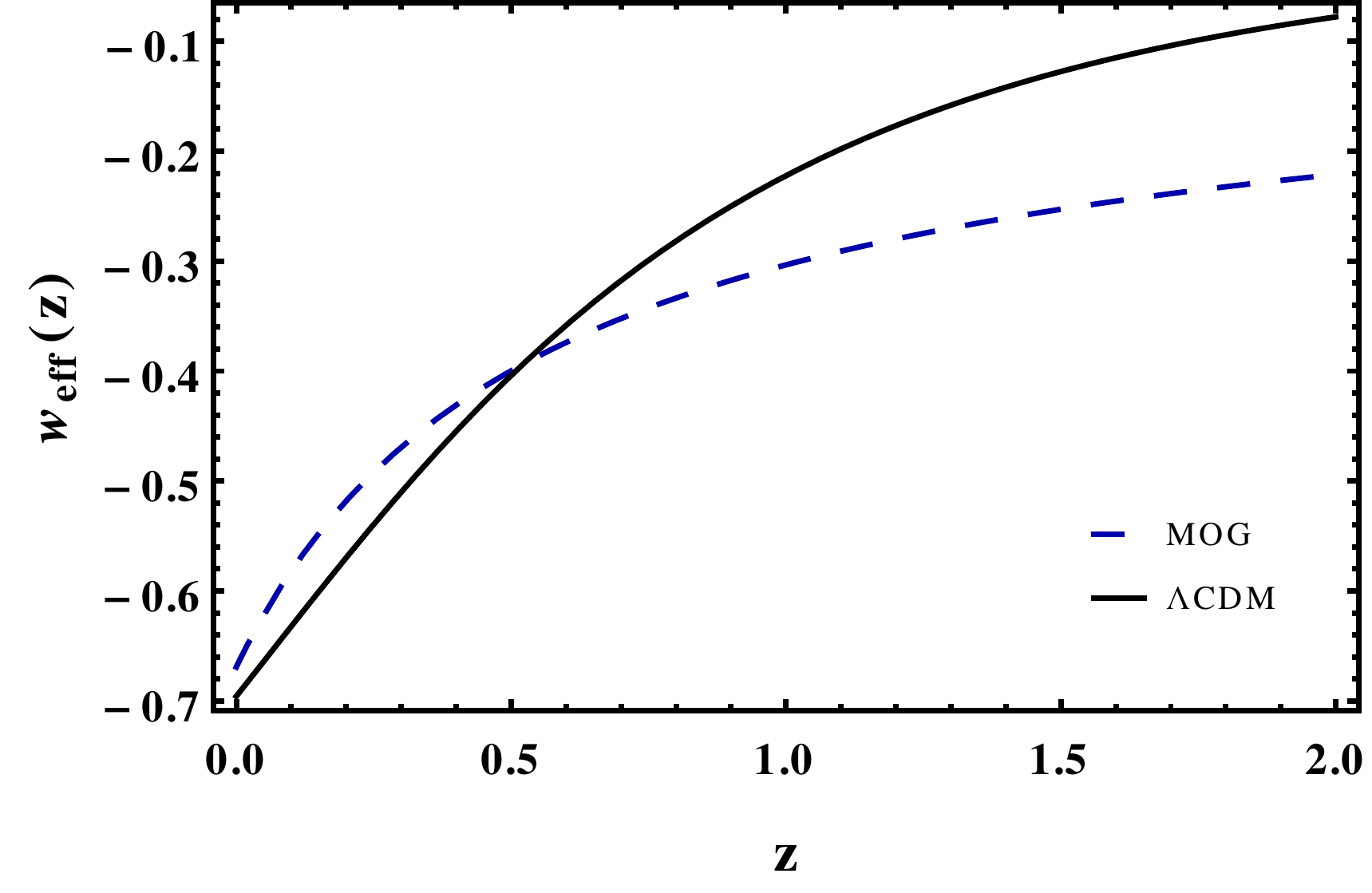}
\caption{The evolution of $w_{\rm eff}(z)$ as a function of the cosmic redshift. See Table~(\ref{tab:res:tot}) for the numerical value of free parameters.} 
		\label{fig-weff-best}
	\end{center}
\end{figure}

\begin{figure}
	\begin{center}
		\includegraphics[width=8cm]{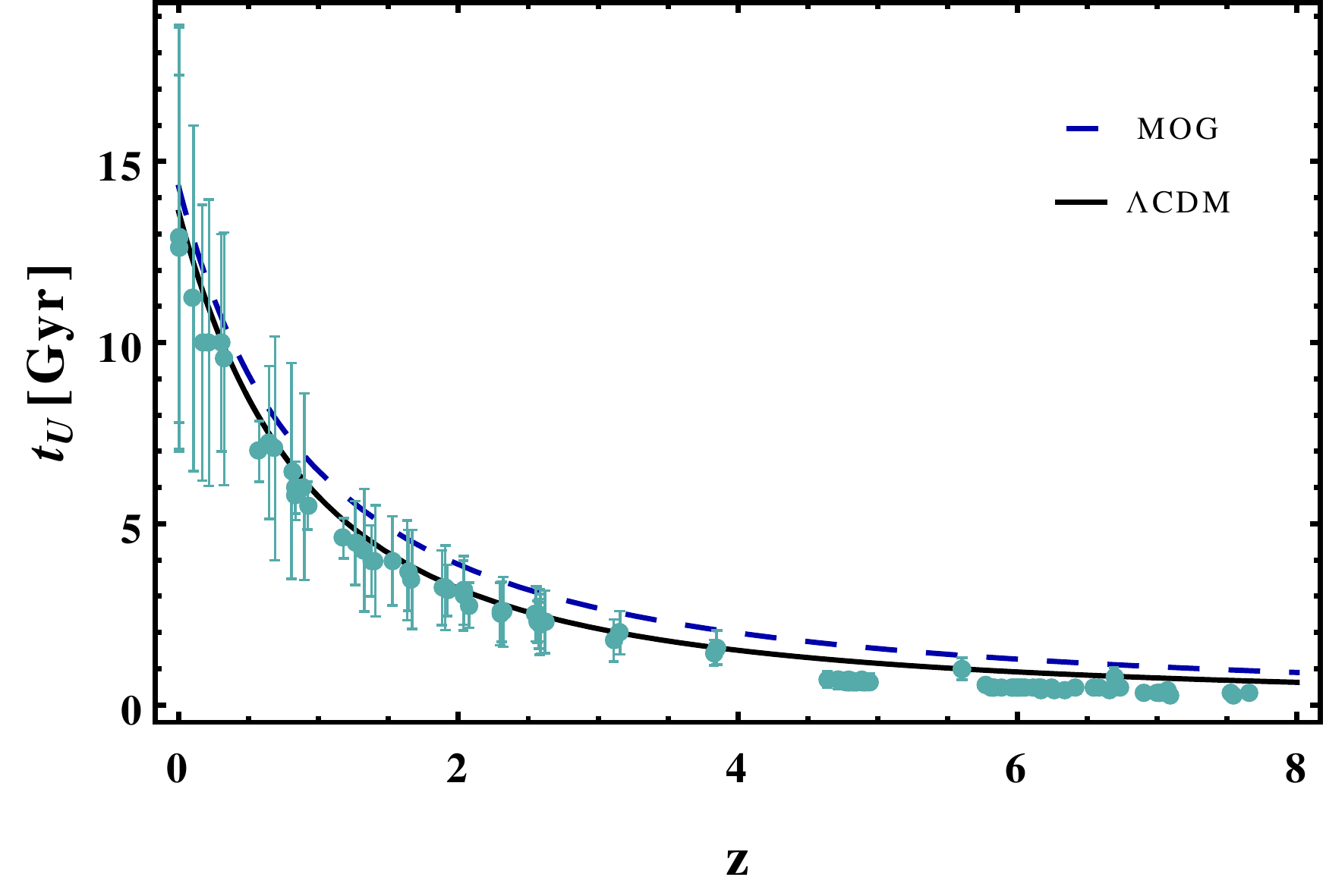}
		\caption{The evolution of the age of the universe, $t_U$ for both MOG theory(the blue dashed curve) and the concordance $\Lambda$CDM (the black solid curve) model as a function of the cosmic
			redshift using best free parameters was reported in Table~(\ref{tab:res:tot}) and the observational data points were reported in~\citep{Vagnozzi:2021tjv} for the ages of old astrophysical objects (OAO).} 
		\label{fig-age}
	\end{center}
\end{figure}

Another parameter we calculated by the best value is the effective EoS and we obtained $w_{\rm eff,\Lambda}=-0.70$ and $w_{\rm eff,MOG}=-0.67$ and in the Figure~(\ref{fig-weff-best}), we show the evolution in term of redshift. 
To this end, using the best-fitting values of Table~(\ref{tab:res:tot}), we can calculate the age of the universe via the following expression
\begin{equation}
t=\int_{0}^{1}\frac{da}{a H(a)}.
\end{equation}
One of the important quantities in cosmology studies is the age of the universe because it connects $H_0$, which can be measured in the late and the early universe. For $\Lambda$CDM, we find $t_{\Lambda}=13.6$ Gyr which are in a good compatibility with the Planck results: 13.78 Gyr;\citep{Aghanim:2018eyx} and for MOG model we obtain a larger value of $t_{\rm MOG}=14.4$ Gyr. In addition to the above equation, the age can also be measured using very old objects. 
Since the mid-1990s, estimates of the ages of globular clusters have consistently been in the range 12-14 Gyr\citep{Bailin:1995dh}. For example, recently it is obtained $t= 13.35 \pm 0.56$ Gyr using
populations of stars in globular clusters\citep{Valcin:2020vav}.  
Despite the robust and rigorous testing that has been done extensively for CMB, BAO, and SN, we need to be more careful about the age of the oldest objects. 
For example, the age of the oldest stars 2MASS J18082002–5104378 B is measured as $t_\star = 13.535\pm0.002$ Gyr, but if the scatter among different models to fit for the age is taken into account
the age becomes $t_\star =13.0 \pm 0.6$ Gyr, and for  HD 140283 is equaled to $t_\star = 14.46 \pm 0.8$ Gyr, but becomes $t_\star = 13.5\pm0.7$ Gyr using the new Gaia parallaxes instead of original HST parallaxes~\citep{DiValentino:2020srs}.\\
The last panel in this study is shown in Figure~(\ref{fig-age}) where we compared the 114 observational data points reported in~\cite{Vagnozzi:2021tjv} for the ages of old astrophysical objects (OAO) in range $z<8$ with theoretical redshift evolution of the age of the universe for both MOG theory and the concordance $\Lambda$CDM model using their best free parameters that were reported in Table~(\ref{tab:res:tot}). As we see two models are consistent with these data.\\
Thus, even if there is no real tension between the different age of the universe determinations at present, most of the error-bars in the age determination comes from the fact that different stellar models do not really agree with each other at the required level of precision to be really able to help with the tensions in cosmology.
\section{CONCLUSIONS}\label{sec:con}	
In this paper, we surveyed the MOG theory as a possible alternative candidate for DE and DM components in cosmology. We performed the present study by a three-step process. First, we solved the system of the basic differential equations (Friedmann and Jeans) at the background and perturbation levels. We investigated the behavior of the basic cosmological quantities $(\Omega(z),H(z),q(z),w_{\rm eff}(z))$ in the background level( see Figures~(\ref{fig1}) \& (\ref{fig2})) and $(f(z),\sigma_8(z),f\sigma_8(z))$ in the perturbation level (see Figure~(\ref{fig3})) in order to understand the global characteristics of the MOG model for different values of its free parameter, $\beta$. As expected, the cosmic expansion depends on the value of $\beta$.\\
Secondly, we performed a combined statistical analysis, involving the
novel geometrical data (BBN, OHD, SNe Type Ia,BAO and CMB shift parameter) and growth data ($f\sigma_8(z)$) and found that the joint and combined statistical
analysis, within the context of flat FLRW space can put sturdy constraints on the basic cosmological parameters.
  As we reported in Table~(\ref{tab:res:tot}), we see that $\Omega_{b0}$ and $\sigma_{80}$ for the MOG model relative to $\Lambda$CDM shift towards a higher value. We found a significant difference between the values of the sound horizon at the baryon drag epoch, $r_{d,fid}$ between MOG and $\Lambda$CDM models.\\
 Thirdly, using the Deviance Information Criterion and the reduced $\chi^2$, we concluded that both MOG and $\Lambda$CDM models fit the observational data well.\\
Lastly, using the aforementioned best cosmological parameters, we found that the age of the universe is greater in the MOG model than in the  $\Lambda$CDM model. Within the context of background dynamics of the universe and growth of structure in the linear regime, MOG is compatible with the observed data.
\section{Acknowledgments}
This work has been supported financially by Iran Science Elites Federation
under research project No. M/99101.
The authors gratefully thank John Moffat and Savvas Nesseris for valuable discussion.
\section{DATA AVAILABILITY}
No new data were generated or analysed in support of this research.
 	\bibliographystyle{mnras}
 	\bibliography{ref}

			\label{lastpage}
\end{document}